\documentclass[useAMS,usenatbib]{mn2e}

\usepackage{epsfig}
\usepackage{graphicx}

\def\ap3m{AP$^3$M}
\def\LCDM{$\Lambda$CDM}

\def\hkpc{$h^{-1}{\ }{\rm kpc}$}
\def\hMpc{$h^{-1}{\ }{\rm Mpc}$}
\def\hMsun{$h^{-1}{\ }{\rm M_{\odot}}$}

\def\nbody{$N$-body}
\def\c15{$c_{\rm 1/5}$}

\def\L30{$\vec{L}_{\rm 30}$}

\def\cosPhi{$\cos \phi$}
\def\Rvir{$R_{\rm vir}$}

\def\vpec{$v_{\rm pec}$}

\def\zform{$z_{\rm form}$}

\def\ea{et~al.~}
\def\lesssim{\mathrel{\hbox{\rlap{\hbox{\lower4pt\hbox{$\sim$}}}\hbox{$<$}}}}
\def\gtrsim{\mathrel{\hbox{\rlap{\hbox{\lower4pt\hbox{$\sim$}}}\hbox{$>$}}}}

\newcommand{\Table}[1]{Table~\ref{#1}}
\newcommand{\Sec}[1]{Section~\ref{#1}}
\newcommand{\Eq}[1]{Eq.~(\ref{#1})}
\newcommand{\Fig}[1]{Figure~\ref{#1}}
\newcommand{\mlapm}{\textsc{mlapm}}
\newcommand{\mhf}{\textsc{mhf}}
\newcommand{\bq}{\begin{equation}}
\newcommand{\eq}{\end{equation}}

\title[The sense of rotation of subhaloes]
{The sense of rotation of subhaloes in cosmological dark matter haloes}

\author[Warnick \& Knebe]
{Kristin Warnick\thanks{E-mail: kwarnick@aip.de} and Alexander Knebe\\
Astrophysical Institute Potsdam, An der Sternwarte 16, D-14482 Potsdam,
Germany\\}

\begin{document}

\date{Accepted 2006 March 22. Received 2006 January 23; in original form 2005 December 5}

\pagerange{\pageref{firstpage}--\pageref{lastpage}} \pubyear{2006}

\maketitle

\label{firstpage}

\begin{abstract}
We present a detailed analysis of the velocity distribution and orientation of orbits of subhaloes in high resolution cosmological simulations of dark matter haloes. We find a trend for substructure to preferentially revolve in the same direction as the sense of rotation of the host halo: there is an excess of prograde satellite haloes. Throughout our suite of nine host haloes (eight cluster sized objects and one galactic halo) there are on average 59\% of the satellites corotating with the host.
Even when including subhaloes out to five virial radii of the host, the signal still remains pointing out the relation of the signal to the infall pattern of subhaloes.
However, the fraction of prograde satellites weakens to about 53\% when observing the data along a (random) line-of-sight and deriving the distributions in a way an observer would infer them. 
This decrease in the observed prograde fraction has its origin in the technique used by the  observer to determine the sense of rotation, which results in a possible misclassification of non-circular orbits.
We conclude that the existence of subhaloes on corotating orbits is another prediction of the cold dark matter structure formation scenario, although there will be difficulties to verify it observationally.
Since the galactic halo simulation gave the same result as the cluster-sized simulations, we assume that the fraction of prograde orbits is independent of the scale of the system, though more galactic simulations would be necessary to confirm this.

\end{abstract}


\section{Introduction}
\label{sec:Introduction}
There is mounting evidence that the Cold Dark Matter (CDM) structure formation scenario provides the most accurate description of our Universe. Observations point towards a ``standard'' \LCDM\ universe comprised of 28\% dark matter, 68\% dark energy, and luminous baryonic matter (i.e. galaxies, stars, gas, and dust) at a mere 4\% \citep[cf. ][]{Spergel03}. This so-called ``concordance model'' induces hierarchical structure formation whereby small objects form first and subsequently merge to form progressively larger objects \citep{WhiteRees78, Davis85, Tormen97}. Whereas the large scale structure of our present Universe can be reconstructed very well by numerical simulations, the small scale structure still poses some problems. For example, there are many more subhaloes predicted by cosmological simulations than actually observed in nearby galaxies \citep[cf. ][]{Klypin99,Moore99}. The lack of observational evidence for these satellites has led to the suggestion that they are completely (or almost completely) dark, with strongly suppressed star formation due to the removal of gas from the small protogalaxies by the ionizing radiation from the first stars and quasars \citep{Bullock00, Tully02, Somerville02}. Others suggest that perhaps low-mass satellites never formed in the predicted numbers in the first place, indicating problems with the \LCDM\ model in general, replacing it with Warm Dark Matter instead \citep{Colin00, Bode01, Knebe02}.  Recent results from (strong) lensing statistics suggest that the predicted excess of substructure is in fact required to reconcile some observations with theory \citep{DalalKochanek02,Dahle03}, although this conclusion has not been universally accepted \citep{SchechterWambsganss02, EvansWitt03, Sand04}. If, however, the lensing detection of halo substructure \textit{is} correct and the overabundant satellite population really does exist, it is vital to understand the orbital evolution of these objects and their deviation from the background dark matter distribution. In order to test the underlying \LCDM\ model, more predictions are necessary which can be confirmed or disproved by observations.

Here, we are investigating cosmological simulations based on the standard $\Lambda$-Cold Dark Matter model, concerning the sense of rotation of subhaloes. 
Host dark matter haloes usually carry a small internal angular momentum,
which is established by the transfer of angular momentum from infalling matter via tidal torques \citep{Peebles69, BarnesEfstathiou87}. However, \citet{Gardner01} as well as \citet{Vitvitska02} proposed another explanation for the origin of the angular momentum in galaxies and their dark matter haloes. They claim that haloes obtain their spin through the cumulative acquisition of angular momentum from satellite accretion. These two descriptions are certainly linked together and mutually dependent, respectively. A detailed analysis of the orbits of satellite haloes shows that they are directly connected to the infall pattern of satellites along the surrounding filaments \citep[e.g.][]{Tormen97,Knebe04,Zentner05}. Those subhaloes falling into the host at early times establish the angular momentum of the inner regions of the primary halo \citep[cf.][]{Vitvitska02} and are channelled into the host along the same direction as those merging at later times.
This leads to the speculation that satellites are preferentially corotating with the host, which is the major assumption (and will be verified in) the current study. 

A similar study was recently presented by \citet{Azzaro05}.
They performed corresponding investigations using both a cosmological simulation and observational data from the Sloan Digital Sky Survey.
These observational data showed a signal of 61\% corotating satellite galaxies, quite independent of magnitudes or distance to the host galaxy. When projecting their simulational data in order to ``observe'' the simulations, they found a fraction of 55--60\% prograde satellite galaxies. 

In another related study, \cite{Shaw05} studied a sample of 2200 (low resolution) dark matter haloes and determined the sense of rotation of all substructure \emph{particles} with respect to the host halo. They found a very strong signal for these particles to be corotating with the host. 
However, we are using here a different method for determining the fraction of prograde satellites, classifying each satellite individually and then counting the number of pro- and retrograde satellites rather than (dark matter) particles.

We present here the analysis of the sense of rotation for subhaloes in nine cosmological simulations: eight cluster-sized objects with varying merger histories and one galactic dark matter halo. These simulations are described in more detail in \Sec{sec:simulations}, while \Sec{sec:simus} deals with the results when analyzing the full six dimensional phase-space information at hand. There we show how the sense of rotation for satellites can be determined and discuss various influences on the fraction of prograde orbits. In \Sec{sec:observer} we investigate how our results can possibly be validated observationally
and conclude with a discussion and summary of our findings in \Sec{sec:discussion} and \ref{sec:conclusions}.

\begin{table*}
\begin{center}
\begin{tabular}{|l|l|c|c|c|c|c|c|c|c|c|c|c|c}
\hline
  host	
& $M_{\rm vir}$
& $R_{\rm vir}$
& $\lambda$
& $V_{\rm max}$
& $R_{\rm max}$
& $\sigma_{v, \rm host}$
& $T$
& age
& $z_{\rm form}$	
& $\left< \frac{\Delta M}{\Delta t M}\right>$
& $\sigma_{\Delta M/M}$
\\ \hline\hline

C1 &   2.9 & 1355 & 0.0157 & 1141 &  346 & 1161 &   0.365 &   7.9  &   1.052& 0.128 &  0.125 \\
C2 &   1.4 & 1067 & 0.0091 &  909 &  338 &  933 &   0.388 &   6.9  &   0.805& 0.122 &  0.156 \\
C3 &   1.1 &  973 & 0.0125 &  828 &  236 &  831 &   0.265 &   6.9  &   0.805& 0.100 &  0.117 \\
C4 &   1.4 & 1061 & 0.0402 &  922 &  165 &  916 &   0.639 &   6.6  &   0.750& 0.127 &  0.207 \\
C5 &   1.2 & 1008 & 0.0093 &  841 &  187 &  848 &   0.909 &   6.0  &   0.643& 0.129 &  0.141 \\
C6 &   1.4 & 1065 & 0.0359 &  870 &  216 &  886 &   0.073 &   5.5  &   0.567& 0.147 &  0.153 \\
C7 &   2.9 & 1347 & 0.0317 & 1089 &  508 & 1182 &   0.531 &   4.6  &   0.443& 0.844 &  1.068 \\
C8 &   3.1 & 1379 & 0.0231 & 1053 &  859 & 1091 &   0.587 &   2.8  &   0.237& 0.250 &  0.225 \\[1ex]
G1 &   1.2$\times 10^{-2}$  &  214 & 0.0229 &  210 &   44 &  202 &  0.491 & 8.5 &   1.232&   0.073 &  0.107 \\ \hline


\end{tabular}
\caption{Properties of the host haloes in our simulations. Masses are measured in 10$^{14}$\hMsun, velocities in km/s, distances in \hkpc, and the age in Gyr. ($M_{\rm vir}$ = virial mass, $R_{\rm vir}$ = virial radius, $\lambda$ = spin parameter, $V_{\rm max}$ = maximum of the rotation curve, $R_{\rm max}$ = position of the maximum, $\sigma_{v, \rm host}$ = velocity dispersion of the host, $T$ = triaxiality parameter, age = time since half of the present day mass has formed, $z_{\rm form}$ = corresponding formation redshift$, \left< \frac{\Delta M}{\Delta t M}\right>$ = mean rate of relative mass change, $\sigma_{\Delta M/M}$ = dispersion of the rate of relative mass change)
}

\label{t:haloprop}
\end{center}
\end{table*}

\section{The Simulations}  \label{sec:simulations}

\subsection{The Raw Data}
Our analysis is based on a suite of nine high-resolution \nbody\ simulations. Eight of them were carried out using the publicly available adaptive mesh refinement code \mlapm\ \citep{mlapm} focusing on the formation and evolution of a dark matter galaxy cluster containing of the order of one million particles, with mass resolution $1.6 \times 10^8$ \hMsun\ and spatial force resolution $\sim$2\hkpc. We first created a set of four independent initial conditions at redshift $z=45$ in a standard \LCDM\ cosmology ($\Omega_0 = 0.3,\Omega_\lambda = 0.7, \Omega_b = 0.04, h = 0.7, \sigma_8 = 0.9$). $512^{3}$ particles were placed in a box of side length 64\hMpc\ giving a mass resolution of $m_p = 1.6 \times 10^{8}$\hMsun.  For each of these initial conditions 
we iteratively collapsed eight adjacent particles to one particle
reducing our particle number to 128$^3$ particles. These lower mass resolution initial conditions were then evolved until $z=0$. At $z=0$, eight clusters (labeled C1-C8) from our simulation suite were selected, with masses in the range  1--3$\times 10^{14}$\hMsun\ and triaxiality parameters \citep[cf. e.g.][]{triaxiality-parameter} from 0.1--0.9 (cf. \Table{t:haloprop}). Then, as described by \citet{Tormen97}, for each cluster the particles within five times the virial radius were tracked back to their Lagrangian positions at the initial redshift ($z=45$). Those particles were then regenerated to their original mass resolution and positions, with the next layer of surrounding large particles regenerated only to one level (i.e. 8 times the original mass resolution), and the remaining particles were left 64 times more massive than the particles resident with the host cluster. This conservative criterion was selected in order to minimise contamination of the final high-resolution haloes with massive particles.

The ninth (re-)simulation was performed using the same (technical) approach but with the {\sc art} (Adaptive Refinement Tree) code \citep{ART-code}. Moreover, this particular run (labeled G1) describes the formation of a Milky Way type dark matter halo in a box of sidelength 20\hMpc. It is the same simulation as ``Box20'' presented in \citet{Box20} and for more details we refer the reader to that study. The final object consists of about two million particles at a mass resolution of $4 \times 10^{7}$\hMsun\ per particle and a spatial force resolution of 0.2\hkpc\ has been reached. All nine simulations have the required resolution to accurately follow the formation and evolution of subhaloes within their respective hosts and hence are well suited for the study presented here.

The host haloes (as well as all substructure objects down to 20 particles) are identified using the open source halo finder \mhf\footnote{\mhf\ (and \mlapm) can be downloaded from the following web page \texttt{http://www.aip.de/People/aknebe/MLAPM}}~\citep[\mlapm's-halo-finder;][]{MHF}. \mhf~is based upon the adaptive grid hierarchy of \mlapm\ and acts with exactly the same 
resolution
as the \nbody\ code itself; it is therefore free of any bias and spurious mismatch between simulation data and halo finding precision arising from numerical effects. For every halo (either host or satellite) we calculate a suite of canonical properties based upon the particles within the virial/truncation radius. The virial radius $R_{\rm vir}$ is defined as the point where the density profile (measured in terms of the cosmological background density $\rho_b$) drops below the virial overdensity $\Delta_{\rm vir}$, i.e. $M(<R_{\rm vir})/(4\pi R_{\rm vir}^3/3) = \Delta_{\rm vir} \rho_b$. This threshold $\Delta_{\rm vir}$ is based upon the dissipationless spherical top-hat collapse model and is a function of 
time for the given cosmological model.
For $z=0$, it amounts to $\Delta_{\rm vir}=340$. This prescription does no longer apply to subhaloes where the point $R_{\rm vir}$ will not be reached due to the embedding of the satellite within the mass distribution of the host, i.e. the density profile will show a rise again at a certain point. In that case, we use this ``upturn point'' and truncate the object ignoring all particles outside of the corresponding sphere. For a more elaborate discussion of this process and the halo finder, in particular, we refer the reader to \citet{Stuart2}.

\subsection{The Host Haloes}
Our halo finder calculates a whole set of integral properties for each individual object, such as virial mass $M_{\rm vir}$, radius $R_{\rm vir}$, spin parameter \citep[cf.][]{spin-parameter}

\begin{equation}
 \lambda = \frac{L}{\sqrt{2} M_{\rm vir} V_{c} R_{\rm vir}} 
\end{equation}

\noindent
(with $L$ = angular momentum, $V_{\rm c} = \sqrt{GM_{\rm vir}/R_{\rm vir}^2}$ = the circular velocity at $R_{\rm vir}$, $G$ = gravitational constant), 
the maximum of the rotation curve $V_{\rm max}$ (and its position $R_{\rm max}$) and the eigenvalues $a>b>c$ of the inertia tensor, which in turn can be used to construct the triaxiality parameter \citep[see e.g.][]{triaxiality-parameter}

\begin{equation}
 T=\frac{a^2-b^2}{a^2-c^2} \ .
\end{equation}

\noindent

In order to get a quantitative measure for the mass accretion history for each of our host haloes, we also compute the dispersion of the fractional mass change rate:

\begin{equation}
 \sigma_{\Delta M/M}^2 = \frac{1}{N_{\rm out}-1} \sum_{i=1}^{N_{\rm out}-1}
                        \left(\frac{\Delta M_i}{\Delta t_i M_i} - 
                         \left< \frac{\Delta M}{\Delta t M}\right> \right)^2 \ ,
\end{equation}

\noindent
where $N_{\rm out}$ is the number of available outputs from formation \zform\ to redshift $z=0$, $\Delta M_i = |M(z_i)-M(z_{i+1})|$ the change in the mass of the host halo, and $\Delta t_i$ is the respective change in time. The formation redshift \zform\ is defined as the redshift where the halo contains half of its present day mass \citep{formation-redshift} and determines the age of the object. The mean growth rate for a given halo is calculated as follows

\begin{equation} \label{MassGrowthRate}
 \left< \frac{\Delta M}{\Delta t M}\right> = 
 \frac{1}{N_{\rm out}-1} \sum_{i=1}^{N_{\rm out}-1} \frac{\Delta M_i}{\Delta t_i M_i} \ .
\end{equation}

\noindent
A large dispersion $\sigma_{\Delta M/M}$ now indicates a violent formation history whereas low values correspond to a more quiescent formation. 
As our definition for formation time implies that the host halo's mass grows by a factor of two until $z=0$, an estimate for the mean growth rate
is simply given by the inverse of the host's age and should be compared against the rate calculated via \Eq{MassGrowthRate}. We summarize these values alongside other relevant properties in \Table{t:haloprop}.

\section{Analyzing the Simulations} \label{sec:simus}

\subsection{Defining the sense of rotation}
In order to determine whether a satellite rotates in the same direction as the host halo (prograde orbit) or in the opposite direction (retrograde orbit), we need to evaluate the alignment of the host's rotation axis and the orbital rotation axis of the satellites.

The rotation axis of objects, such as host dark matter haloes, is not well defined\footnote{For a rigid body, the relation between the angular velocity vector $\vec{\omega}$ determining the rotation axis and the angular momentum vector $\vec{L}$ can be derived via $\vec{L} = I \vec{\omega}$, where $I$ is the inertia tensor.}, thus we use the angular momentum for the description of its rotation:

\begin{equation}
 \vec{L}(R) = \sum_{r_i<R} m_i \vec{r}_i \times \vec{v}_i   \ ,
\end{equation}

\noindent
which is being calculated at various distances $R$ from the host's centre using all interior particles. The (dark matter particle) velocities $\vec{v}_i$ here are in the rest frame of the host as are the particles' positions $\vec{r}_i$.

In order to eliminate the influence of (massive) substructures at large halocentric distances we used the angular momentum vector \L30\ as defined by the material inside a sphere containing 30\% of the host's virial mass ($\approx 0.2$\,\Rvir\ on average). This also ensures to focus on the properties of the inner halo whose material collapsed first \citep[cf.][]{innerhalomaterialcollapse}. 

The rotation axis of a satellite orbit simply coincides with its orbital angular momentum vector 

\begin{equation}
 \vec{L}_{\rm sat} = m_{\rm sat} \vec{r}_{\rm sat} \times \vec{v}_{\rm sat} \ .
\end{equation}

\noindent
where $\vec{r}_{\rm sat}$ and $\vec{v}_{\rm sat}$ are again measured in the rest frame of the host halo. 

To check whether a satellite is co- or counterrotating we use the angle between the internal angular momentum of the host and the orbital angular momentum of the satellite, i.e. more precisely we use the scalar vector product

\begin{equation} \label{L30Lsat}
 \cos \phi = \frac{\vec{L}_{\rm 30} \cdot \vec{L}_{\rm sat}^{\rm orb}}{{L}_{\rm 30} {L}_{\rm sat}^{\rm orb}}
\end{equation}

\noindent
and define a satellite to be prograde for $\cos\phi > 0$ and retrograde for $\cos\phi < 0$.
Thus, we also include satellites on nearly ``perpendicular'' orbits ($\cos\phi \approx 0$), whose classification as either pro- or retrograde may be questionable. However, such subhaloes are part of our sample and any hypothetical observer of the system will have to deal with them, too. We will come back to this point in \Sec{sec:observer}, where we project our data into the ``observer's plane''.

\subsection{Determining the sense of rotation}
\begin{figure}
\begin{center}
	\begin{tabular}{p{25mm}p{25mm}p{25mm}}
	   	\resizebox{27mm}{!}{\epsfig{file=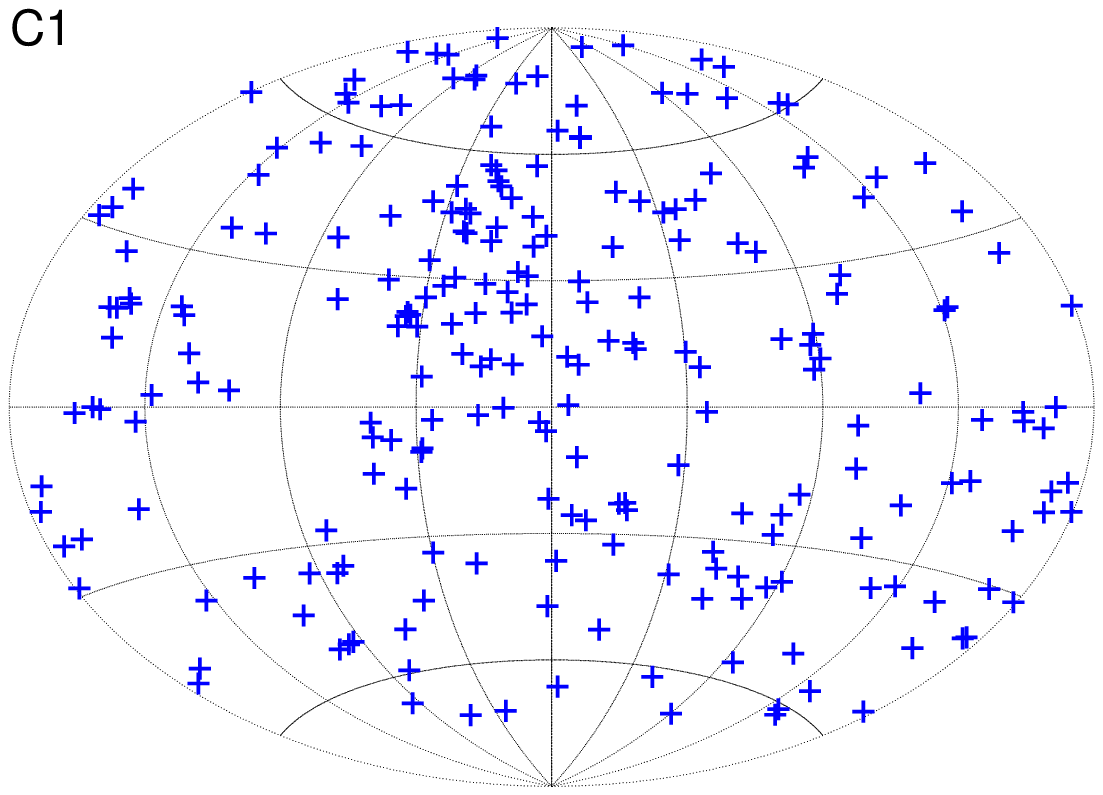, angle=-90}} &
		\resizebox{27mm}{!}{\epsfig{file=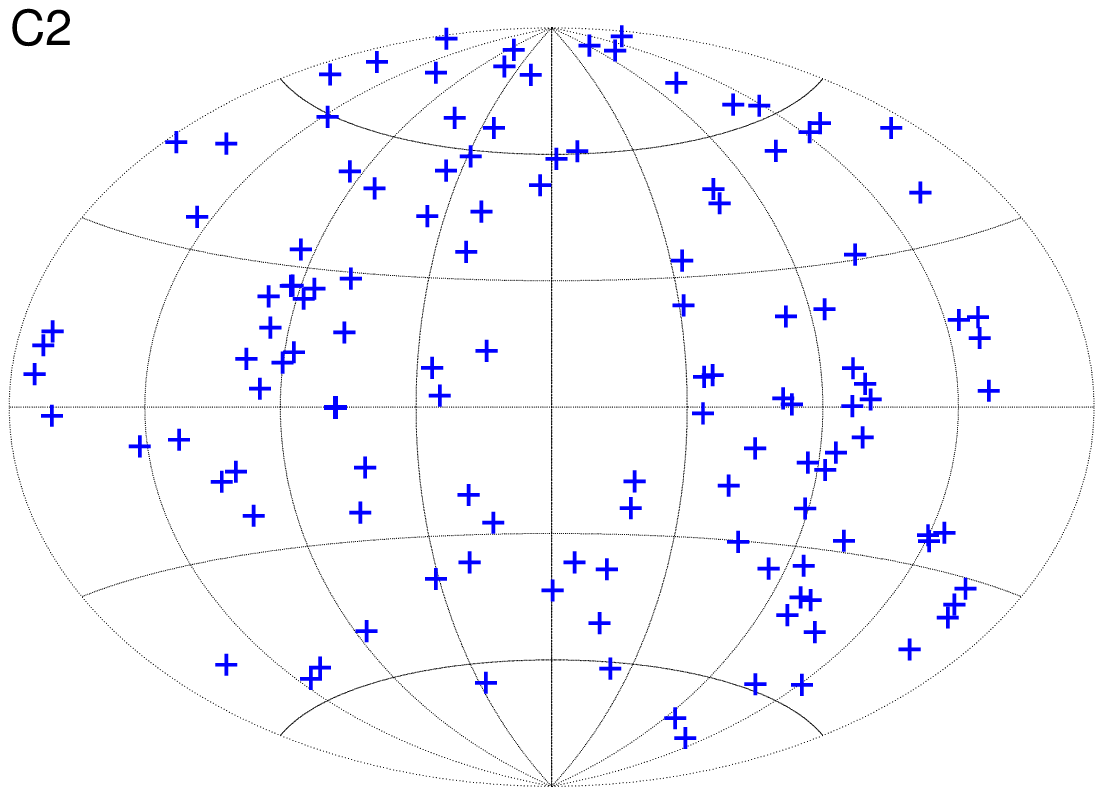, angle=-90}} &
		\resizebox{27mm}{!}{\epsfig{file=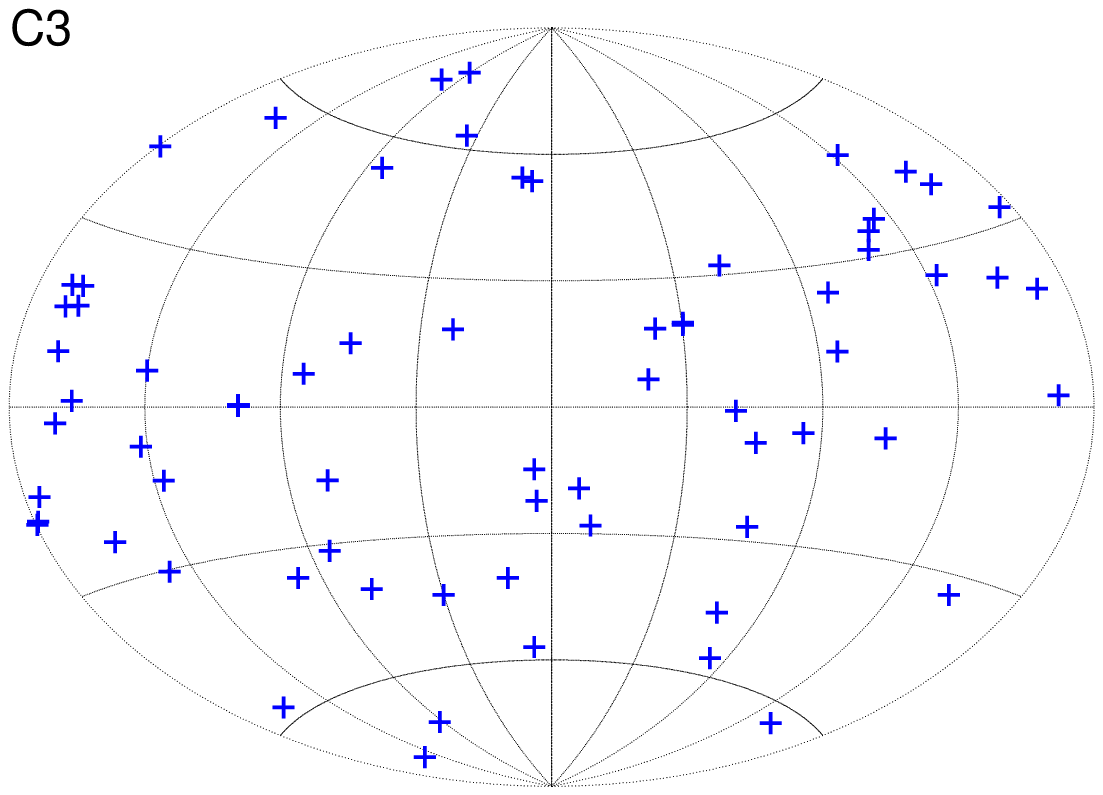, angle=-90}} \\
	   	\resizebox{27mm}{!}{\epsfig{file=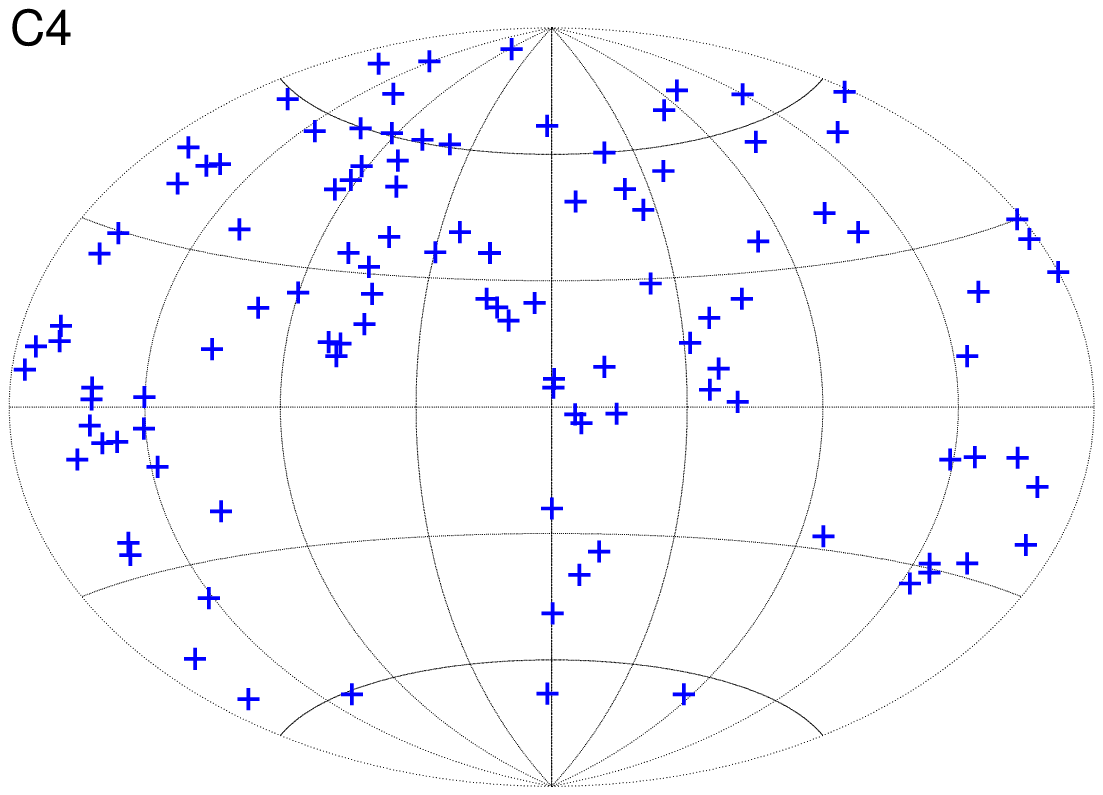, angle=-90}} &
		\resizebox{27mm}{!}{\epsfig{file=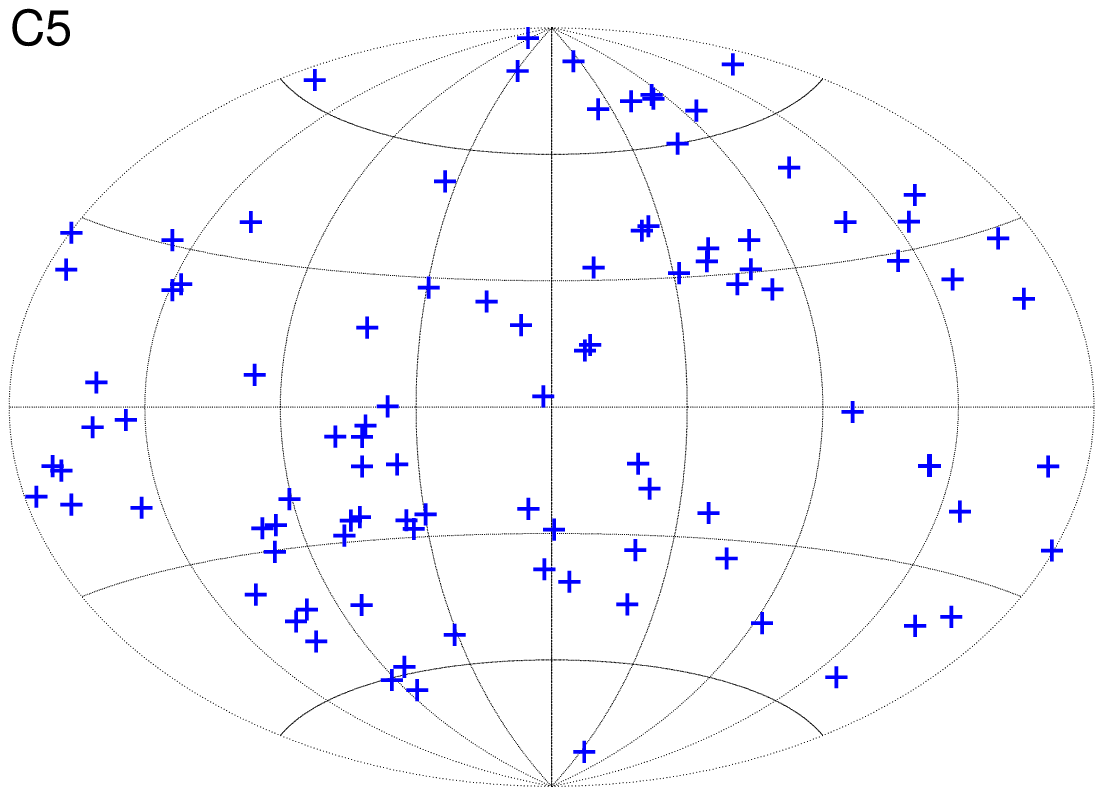, angle=-90}} &
		\resizebox{27mm}{!}{\epsfig{file=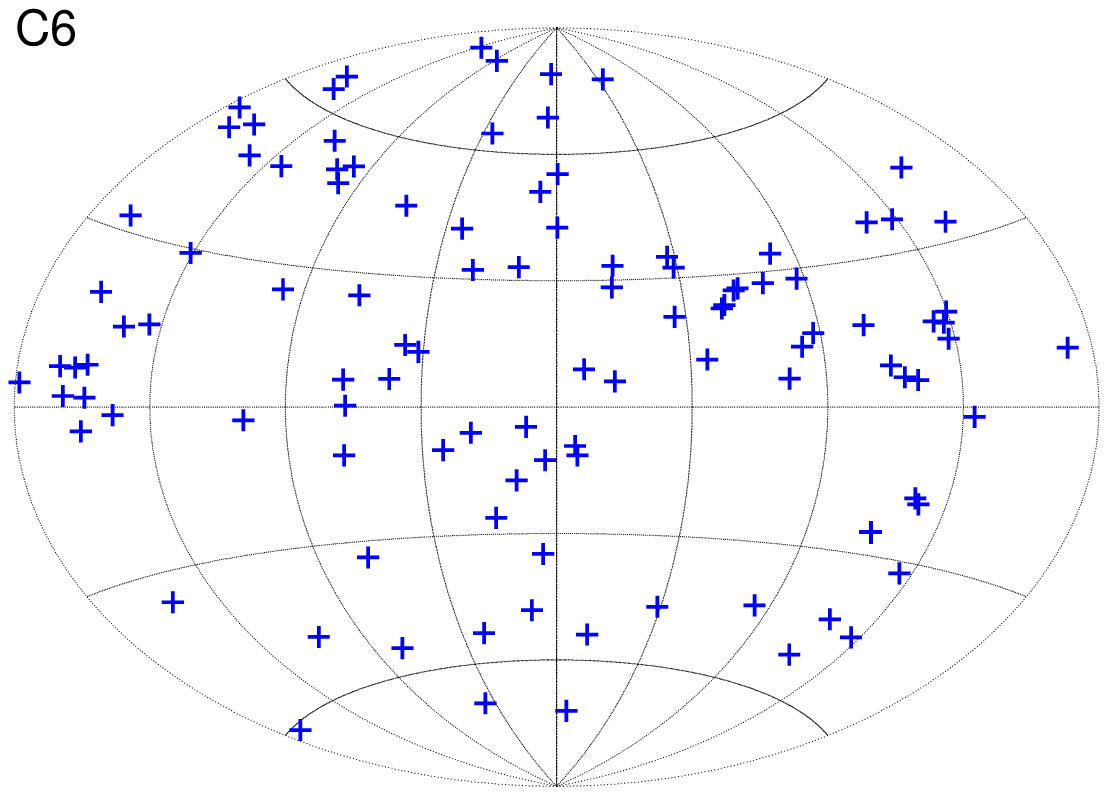, angle=-90}} \\
	   	\resizebox{27mm}{!}{\epsfig{file=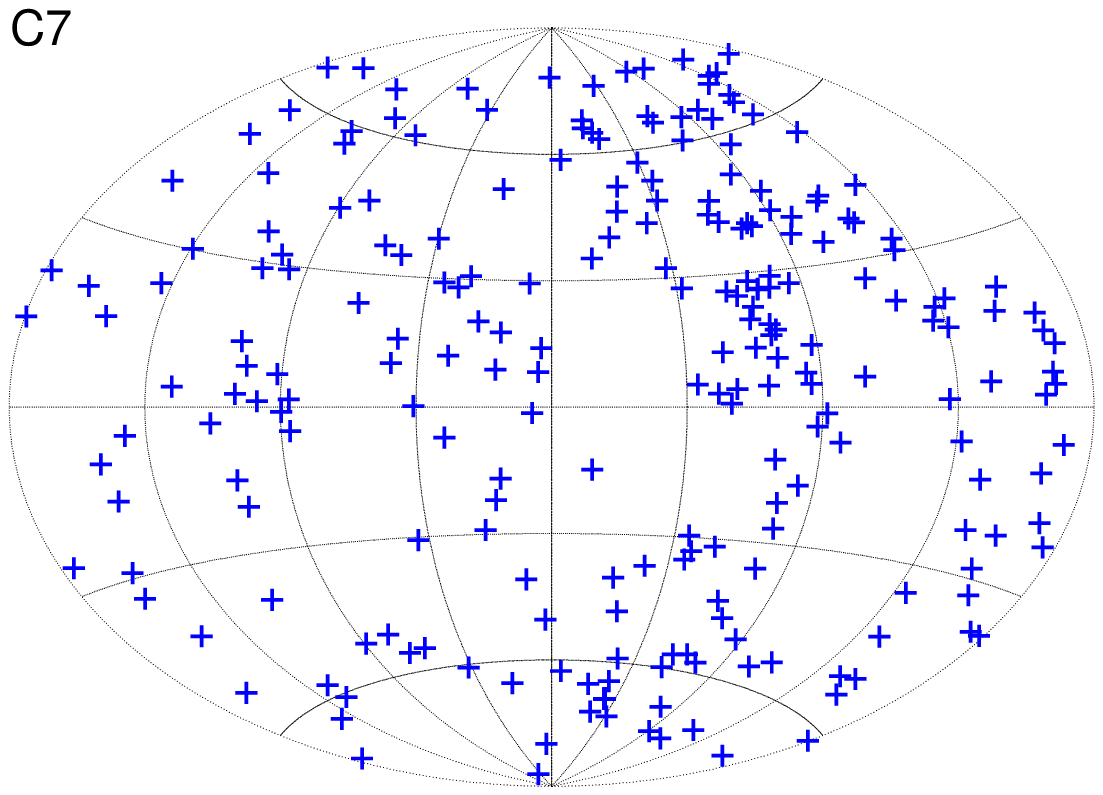, angle=-90}} &
		\resizebox{27mm}{!}{\epsfig{file=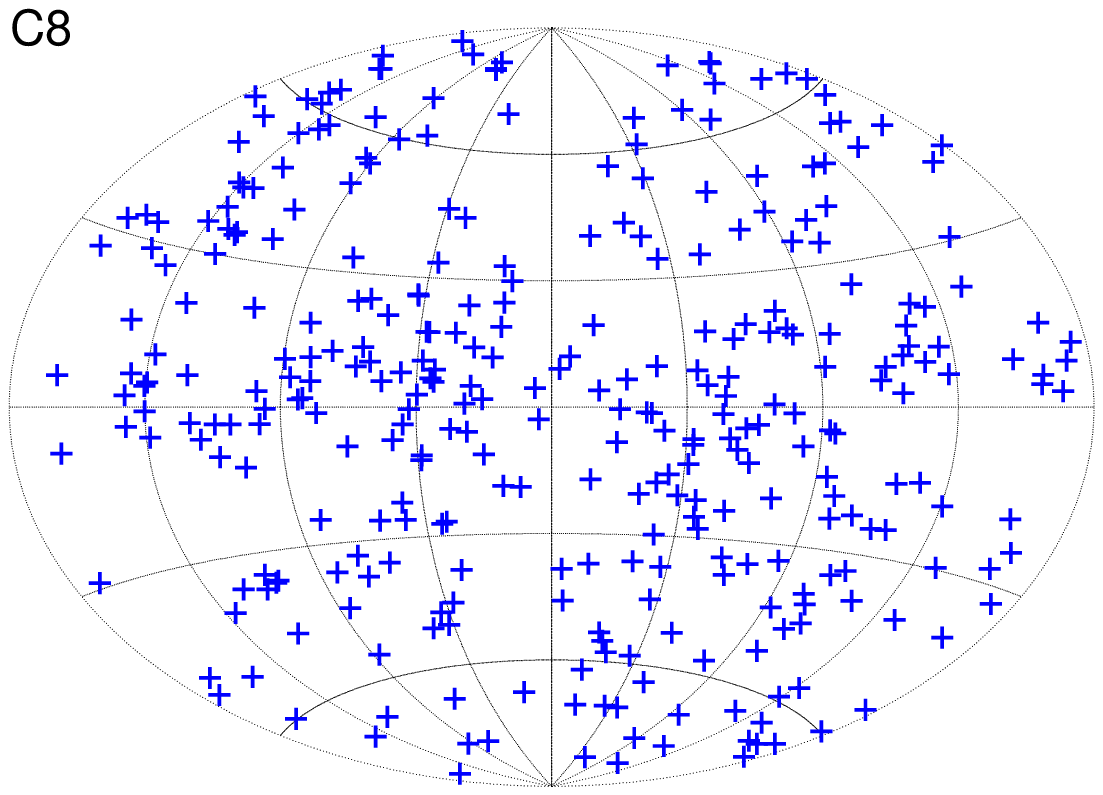, angle=-90}} &
		\resizebox{27mm}{!}{\epsfig{file=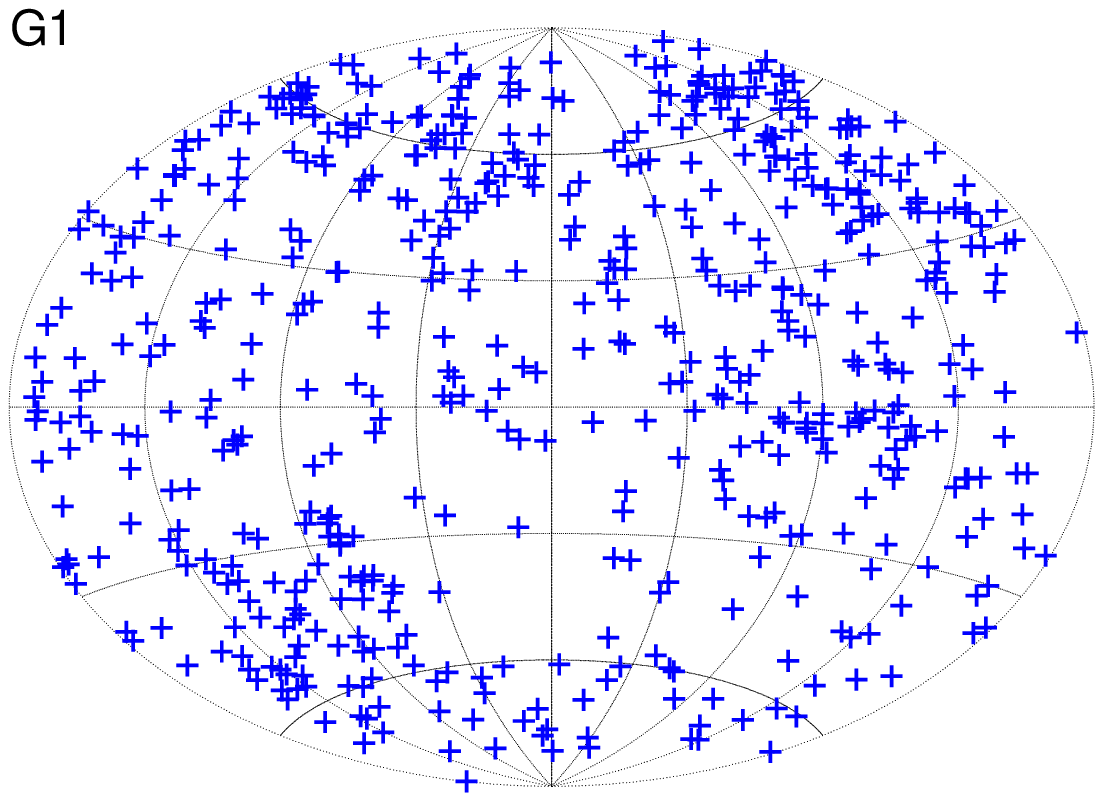, angle=-90}} \\
	\end{tabular}\\[0.65cm]
\end{center}
\caption
{The distribution of directions of orbital angular momenta on a unit sphere. The host's internal angular momentum points from the centre to the North Pole. The points where the orbital angular momenta of satellites pierce through the unit sphere are marked with a '+'.
More points in the Northern hemisphere than in the Southern hemisphere indicate a majority of prograde orbits.}
\label{f:mapsat}
\end{figure}

\begin{figure*}
\begin{center}
	\epsfig{file=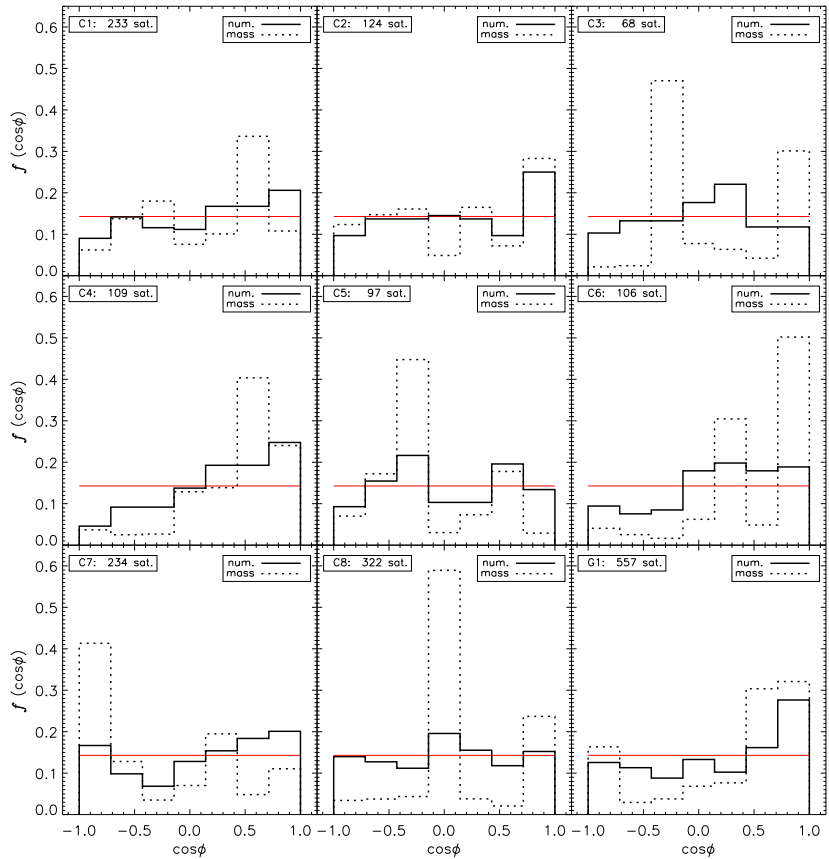, height=0.7\textwidth, angle=0}
\end{center}
\caption{The distribution of the cosines of angles $\cos\phi$ between the orbital angular momentum of the satellites and the internal angular momentum of the host halo for nine different simulations (first panel = oldest halo) at redshift $z=0.0$. Shown is the fraction of the number of satellites per $\cos\phi$-bin (black solid line), divided by the size of each bin. The red horizontal line marks the value for isotropic distribution. The dotted curve represents the corresponding mass fraction of the satellites per bin.}
\label{f:orientsatbin_both}
\end{figure*}

\begin{figure*}
\begin{center}
	\epsfig{file=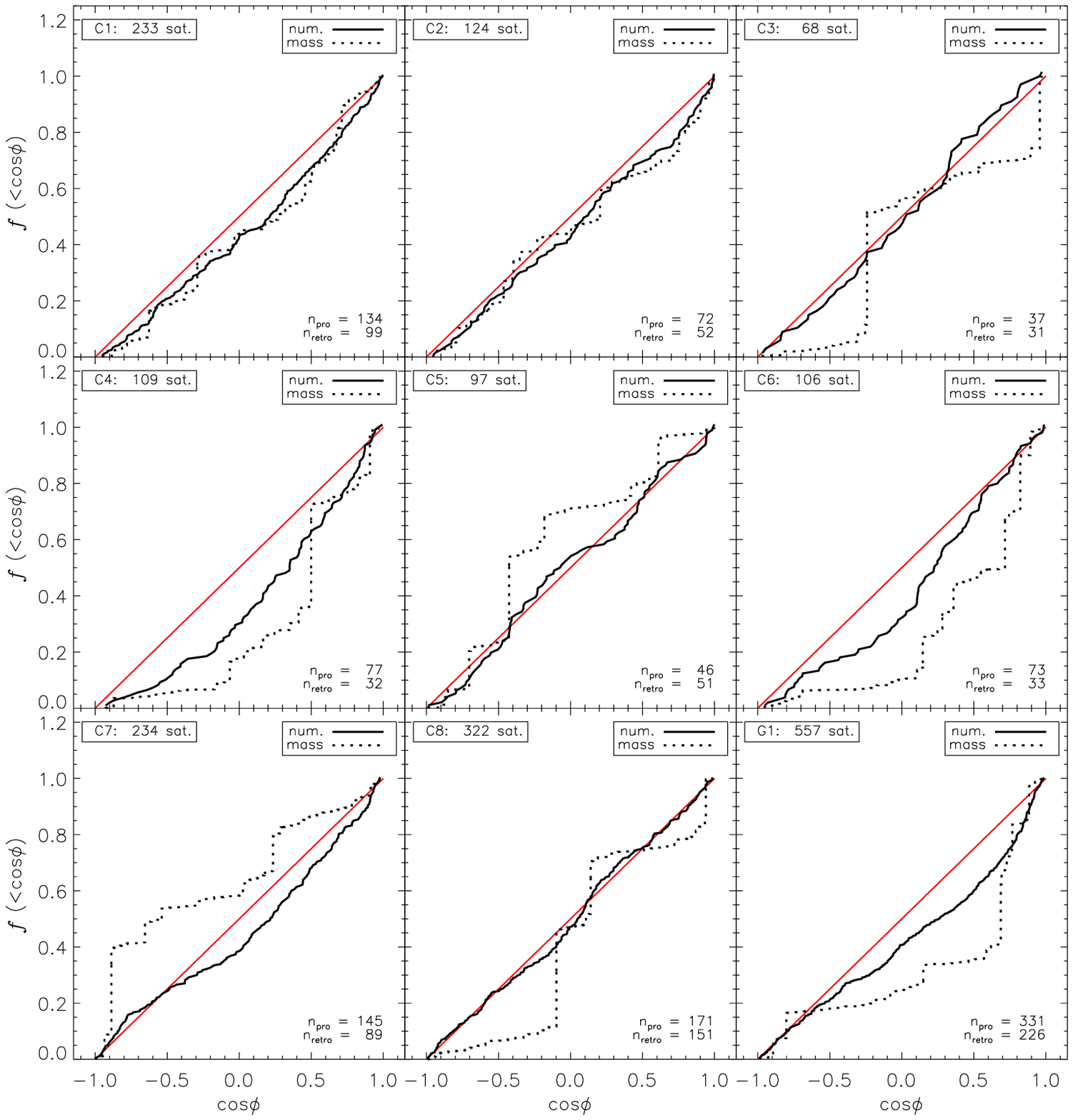, height=0.7\textwidth, angle=0}
\end{center}
\caption
{The cumulative distribution of the cosines of angles $\cos\phi$ between the orbital angular momentum of the
satellites and the spin of the host halo for nine different simulations (first panel = oldest halo) at redshift z=0.0 (black solid line). The dotted line shows the cumulative distribution with mass weighing -- instead of just counting the satellites, the masses of the satellites were added. A strong increase in that curve indicates a massive satellite. The red line shows the ideal case of an isotropic distribution. 
}
\label{f:orientsatcum_both}
\end{figure*}

For a visual impression of the distributions of orientations of subhaloes and their (an-)isotropy, we plot the direction of the orbital angular momenta on a unit sphere in \Fig{f:mapsat} (projected to a plane using the Aitoff-Projection). The panel for C4 clearly shows a majority of points in the Northern hemisphere, indicating an excess of prograde orbits. In contrast, the distributions for the hosts C3, C5 and C8 appear highly isotropic. A quantitative analysis of these first impressions will follow in the next sections.

\subsubsection{Differential distribution of angles}
We determined the cosine of the angle between the inner host halo angular momentum $\vec{L}_{\rm 30}$ and the orbital angular momentum $\vec{L}_{\rm sat}^{\rm orb}$ for every satellite residing within the virial radius \Rvir\ of the host halo. \Fig{f:orientsatbin_both} now shows the (binned) differential distribution of \cosPhi. We readily observe a (slight) increase in the fraction of satellites for larger values of $\cos\phi$, indicating an overbalance of prograde orbits. While the solid lines represent the distributions when counting each satellite once (``number weighing''), the dotted line is derived by weighing the contribution of each satellite by its mass. Comparing both curves demonstrates the importance of high-mass satellites on the prograde signal. Unfortunately the trend is not that obvious and we rather conclude that the fraction of prograde subhaloes is not significantly 
influenced by 
heavier satellites. The red solid horizontal line in each plot marks the expected distribution for an isotropic distribution of \cosPhi.

\subsubsection{Cumulative distribution}
Presenting a differential distribution always includes the problem of binning the data and hence introducing (unnecessary) noise, especially for small number statistics. To avoid this and provide a smoother view of the data, we examine the cumulative distribution function, as shown in \Fig{f:orientsatcum_both}. Here we present the number fraction (and the mass fraction, dotted line) of satellites with \cosPhi\ smaller than a given value.  For an isotropic distribution we expect a straight diagonal line as indicated by the red solid line. 
We can see, that the distributions in \Fig{f:orientsatcum_both} do not seem to be isotropic, just as already indicated by \Fig{f:orientsatbin_both}. However, this does not necessarily lead to a bias towards prograde orbits. If we still had the same number of pro- and retrograde orbits, the distributions would be point symmetric. But we observe the clear trend in (most of) our systems that the data curves lie below the isotropic expectation.

%
\begin{table}
\begin{center}
\begin{tabular}{cclrrrrc}
\hline
host & $D$ & $\hphantom{01}p$ & $\cos\Phi_{\rm net}$ & $n_{\rm sat} $ & $n_{p}$ & $n_{r}$ & $f_{p}$ \\ \hline
C1 & 0.120 & 0.064 & 0.959 & 233 & 134 & 99 & 0.575 \\
C2 & 0.113 & 0.387 & 0.318 & 124 & 72 & 52 & 0.581 \\
C3 & 0.073 & 0.991 & 0.248 & 68  & 37 & 31 & 0.544 \\
C4 & 0.220 & 0.008 & 0.897 & 109 & 77 & 32 & 0.706 \\
C5 & 0.062 & 0.990 & $-0.286$ & 97 & 46 & 51 & 0.474 \\
C6 & 0.198 & 0.027 & 0.978 & 106 & 73 & 33 & 0.689 \\
C7 & 0.124 & 0.050 & 0.369 & 234 & 145 & 89 & 0.620 \\
C8 & 0.059 & 0.617 & 0.206 & 322 & 171 & 151 & 0.531 \\[1ex]
G1 & 0.156 & 10$^{-6}$ & 0.622 & 557 & 331 & 226 & 0.594 \\ \hline\hline
\multicolumn{8}{r}{$<f_p>$ = $0.590 \pm 0.070$}  \\ \hline\hline
\end{tabular}
\caption{Isotropy and prograde fraction. The isotropy was checked with a KS test, which calculates the maximum deviation $D$ and the probability $p$ for the consistency with an isotropic distribution. $\cos\Phi_{\rm net}$ is the cosine of the angle between the inner host angular momentum and the net angular momentum of all satellite orbits. $n_{\rm sat}$ is the total number of satellites that can be split into $n_{p}$ and $n_r$, the number of prograde and retrograde orbits, respectively. The prograde fraction $f_p$ is given in the last column and the last line simply gives the mean fraction (along with its standard deviation) of prograde orbits when averaging over all hosts.}
\label{t:kstestnum}
\end{center}
\end{table}

\subsubsection{Confirming the signal via a Kolmogorov-Smirnov test} \label{sec:KStest}
In order to validate the preference for satellites to be on prograde orbits as seen in both Figs \ref{f:orientsatbin_both} and \ref{f:orientsatcum_both} and to quantify the significance of the signal, we carried out a Kolmogorov-Smirnov (KS) test \citep[e.g.][]{NumRec} for the number weighted distributions. The KS test provides us with the maximum deviation of the data set from the isotropic distribution ($D$) and the probability ($p$) for the data being consistent with isotropy. These numbers are listed in \Table{t:kstestnum} and show that only the distributions of orientations in simulations C3 and C5 (and possibly C8) are close to isotropy with a nearly identical number of pro- ($n_p$) and retrograde ($n_r$) orbits. The latter numbers are also given in \Table{t:kstestnum} supplemented by the fraction of subhaloes on prograde orbits, i.e. $f_p=n_p/(n_p+n_r)$. In all other simulations we find a quantitative confirmation of an excess of prograde orbits.

However, as mentioned above, a deviation from isotropy does not automatically mean an overbalance of prograde orbits: it is the asymmetry of the distribution, which could validate the result.
If the satellites were equally distributed between pro- and retrograde orbits, the differential distribution would be symmetric about the vertical line $\cos\phi = 0$ and the cumulative distribution function would be point symmetric around \mbox{(0, $f(<0.5)$)}. We can test for this (point) symmetry by plotting the cumulative distribution not in the interval $[-1,+1]$ but rather in reverse with \cosPhi\ ranging from $[+1,-1]$.   If these two ways of calculating the distribution give rise to identical curves, then the distribution is (point) symmetric, without any preference for pro- or retrograde orbits. In order to further quantify this kind of symmetry, we can again apply a KS test to these two ``mirror''-distributions. For six of our nine simulations we could not detect any point symmetry, the probability for point symmetry is at most 5\%. Only for the simulations C3 (84.4\%), C5 (88.5\%) and C8 (19.8\%) we do find a stronger sign of symmetry. These systems are the cases, which also show more or less isotropic distributions and thus this result comes at no surprise as isotropy entails (point) symmetry. We will elaborate upon these three systems later on in \Sec{discussion1}. However, for all the other simulations, this analysis confirms the credibility of our results which can be summarized by the mean fraction of prograde orbits of \mbox{59\% $\pm$ 7\%} throughout all simulations. 

%

The strength of the signal can be measured not only by the (number) fraction of prograde orbits as given in the last column of \Table{t:kstestnum} but also by (the cosine of) the angle between the host's angular momentum and the net angular momentum of all subhaloes

\begin{equation}
 \cos \Phi_{\rm net} = \frac{\vec{L}_{\rm 30} \cdot \sum_{i=1}^{n_{\rm sat}} \vec{L}_{{\rm sat},i}^{\rm orb}}{{L}_{\rm 30}  |\sum_{i=1}^{n_{\rm sat}} \vec{L}_{{\rm sat},i}^{\rm orb}|} \ .
\end{equation}

\noindent
These values are also listed in \Table{t:kstestnum} and with one exception (C5, the most isotropic case) we only have positive values, stressing again the preference of prograde orbits. One needs to bear in mind though that the use of $\cos \Phi_{\rm net}$ rather confirms a ``mass weighted signal'' as the angular momentum $\vec{L}_{{\rm sat},i}^{\rm orb}$ depends on the mass of each individual subhalo.



\subsubsection{Uncertainties in the host's rotation axis} \label{sec:SpinCheck}
For the analysis so far, we used the angular momentum of the inner halo, i.e. \L30 as defined by the material inside a sphere encompassing 30\% of the virial mass. We now examine the effect when using the angular momentum based upon larger fractions, i.e. 50\% or even 100\%, of the host's virial mass. 

We expect variations in the host's angular momentum for two reasons: first, a dark matter halo is \emph{not} a rigid body and secondly, the angular momentum in the outer parts may be ``contaminated'' by recently accreted material and infalling satellites. These variations in the direction of the host angular momentum vector compared to the angular momentum of the inner 30\% are quantified in \Table{t:angles} where we list the angle between the two respective vectors

\begin{equation}
 \alpha^{\rm 50/100} = \cos^{-1} \left( \frac{\vec{L}_{\rm 30} \cdot \vec{L}_{\rm 50/100}}{{L}_{\rm 30} {L}_{\rm 50/100}} \right) \ .
\end{equation}

\noindent
The last line in \Table{t:angles} summarizes the mean angle when averaging over all hosts along with its standard deviation. 

\begin{table}
\begin{center}
\begin{tabular}{ccccc}
\hline
host & $\alpha^{50}$ [$^\circ$] & $\alpha^{100}$ [$^\circ$]\\ \hline \hline
C1 & 32.6  & 27.0\\
C2 & 19.1  & 46.5\\
C3 & 28.4  & 35.4\\
C4 & \hphantom{0}2.4  &  \hphantom{0}7.4\\
C5 & 48.8  & 37.3\\
C6 & \hphantom{0}3.4  & 17.4\\
C7 & 40.5  & 43.0\\
C8 & 16.9  & 65.2\\[1ex]
G1 & 15.3  & 32.8\\ \hline \hline
$<\alpha>$ & $23 \pm 15$ & $35 \pm 17$\\ \hline \hline

\end{tabular}
\caption{The difference in the direction of the host angular momentum at radii including 50\% and 100\% of the host mass, with respect to the 30\%-angular momentum.}
\label{t:angles}
\end{center}
\end{table}

More interesting now are the variations in the fraction of prograde orbits, i.e. $\Delta f_p= f_p-f_p^{30}$. \Table{t:profrac-angmom} summarizes the corresponding prograde fractions ($f_p^{50}$ and $f_p^{100}$) and their deviations $\Delta f_p= f_p-f_p^{30}$ from our reference ``inner'' fractions $f_p^{30}$. Even though there are minor variations amongst individual haloes, the mean fraction of prograde orbits stays rather constant irrespective of the definition for the host's angular momentum. 

This result can be understood considering the fact that there are not many satellites on perpendicular orbits as indicated by \Fig{f:mapsat} -- the orbital angular momenta preferentially point to the region around the poles rather than to the equator region. Since we define all satellites as corotating whose angular momenta lie north of the equator, variations in the direction of the angular momenta do not have a prominent influence on the number of prograde orbits.

Hence, despite
the rather large differences in the relative orientation of the respective angular momenta (cf. \Table{t:angles}) the influence of the point, where to define the host's angular momentum, on the mean fraction of prograde orbits is only minor.

\begin{table}
\begin{center}
\begin{tabular}{ccccc}
\hline
host & $f_p^{50}$ & ($\Delta f_p^{50}$) & $f_p^{100}$ & ($\Delta f_p^{100}$) \\ \hline\hline
C1 & 0.601 & $(+0.026)$ & 0.579 & $(+0.004)$ \\
C2 & 0.556 & $(-0.024)$ & 0.605 & $(+0.024)$ \\
C3 & 0.471 & $(-0.074)$ & 0.515 & $(-0.029)$ \\
C4 & 0.706 & $(+0.000)$ & 0.725 & $(+0.018)$ \\
C5 & 0.588 & $(+0.113)$ & 0.660 & $(+0.186)$ \\
C6 & 0.689 & $(+0.000)$ & 0.585 & $(-0.103)$ \\
C7 & 0.585 & $(-0.034)$ & 0.546 & $(-0.073)$ \\
C8 & 0.481 & $(-0.050)$ & 0.488 & $(-0.043)$ \\[1ex]
G1 & 0.639 & $(+0.045)$ & 0.627 & $(+0.032)$ \\ \hline\hline
$<f_p>$ & \multicolumn{2}{c}{$0.591 \pm 0.077$} & \multicolumn{2}{c}{$0.592 \pm 0.069$} \\ \hline\hline

\end{tabular}
\caption{Quantifying the influence of the host's angular momentum (at 50\% and 100\% of the host mass) on the fraction of prograde orbits. The last line gives the mean prograde fractions with standard deviations.}
\label{t:profrac-angmom}
\end{center}
\end{table}

%

%
\begin{table}
\begin{center}
\begin{tabular}{ccccccc}
\hline
   & \multicolumn{2}{c}{$1.5 \cdot R_{\rm vir}$} 
	& \multicolumn{2}{c}{$3.0\cdot R_{\rm vir}$} 
	& \multicolumn{2}{c}{$5.0\cdot R_{\rm vir}$} \\
host & $n_{\rm sat}$ & $f_p$   & $n_{\rm sat}$ & $f_p$  & $n_{\rm sat}$ & $f_p$\\ \hline\hline
C1 & 458 & 0.544 & 750 & 0.507 & 774 & 0.509 \\
C2 & 243 & 0.560 & 384 & 0.534 & 411 & 0.530 \\
C3 & 133 & 0.534 & 192 & 0.599 & 201 & 0.602 \\
C4 & 212 & 0.693 & 311 & 0.695 & 318 & 0.686 \\
C5 & 144 & 0.472 & 210 & 0.471 & 226 & 0.469 \\
C6 & 199 & 0.638 & 365 & 0.592 & 395 & 0.587 \\
C7 & 382 & 0.613 & 632 & 0.568 & 659 & 0.561 \\
C8 & 446 & 0.545 & 570 & 0.549 & 604 & 0.558 \\[1ex]
G1 & 904 & 0.579 & 1589 & 0.566 & 3015 & 0.512 \\ \hline\hline
$<f_p>$ 
     & \multicolumn{2}{c}{$0.575 \pm 0.061$}
     & \multicolumn{2}{c}{$0.565 \pm 0.060$}
     & \multicolumn{2}{c}{$0.557 \pm 0.060$}  \\ \hline\hline

\end{tabular}
\caption{Numbers of satellites and prograde fractions when including satellites from the immediate environment, i.e. including satellites outside the virial radius of the host halo, up to 1.5, 3.0 and $5.0 R_{\rm vir}$. The mean values of the prograde fractions $<f_p>$ for all simulations are provided in the last line.
}
\label{t:profrac-environment}
\end{center}
\end{table}
%

\subsubsection{Environmental dependence} \label{sec:environment}
The recent study of \citet{Box20} showed that isolated galaxy-sized haloes display all the properties of relaxed objects up to 2--3\,\Rvir. One of their haloes under investigation (i.e. ``Box20'') was in fact the galactic halo also used in our study. Further, \citet{backsplash-galaxies} showed that there exists a prominent population of backsplash subhaloes: satellites that once passed the host's virial radius but are now residing in the outskirts of the halo. It therefore appears interesting to investigate the effect of allowing for satellites outside the virial radius to be included in the determination of the prograde fraction. The results for gradually increasing the trading area and considering satellites within a sphere of radius 1.5, 3, and 5\,\Rvir\ are given in \Table{t:profrac-environment}. Besides listing the updated prograde fraction we also present the number of satellites in the respective sphere. We can see that while the number of subhaloes grows substantially, the sense of orientation of the satellites is hardly affected. The signal weakens becoming more isotropic, but yet remains even at a distance of 5\,\Rvir.

\subsection{Discussion} \label{discussion1}
We have seen in our simulations that the average fraction of prograde orbits is $0.590 \pm 0.070$.

In \Sec{sec:Introduction} we sketched a scenario, where matter streams into the (progenitors of) host haloes from preferred directions, i.e. along the surrounding filaments, thus establishing the angular momentum of the host. Since today's satellites are also expected to fall into the host from the same preferred directions, this picture suggests an overabundance of prograde orbits. This scenario is supported by the analysis presented in the previous \Sec{sec:environment}. Even satellites as far away from the host as 5\,\Rvir\ will eventually fall into the host in a way that complies with the already established sense of rotation.

Though this picture in general seems to be confirmed, we still would like to understand why not all our systems do exhibit the signal.
If we use the information about the hosts presented in \Table{t:haloprop} and compare their integral properties with the prograde fraction for each halo, we note a slight trend of larger prograde fractions with increasing spin parameter (neglecting the galactic halo for the moment). The isotropic hosts C3 and C5 possess only a very small spin parameter of $\lambda = 0.0125$ and 0.0093, respectively. This connection is understandable when considering the origin of the host angular momentum from mergers with subhaloes -- the smaller the angular momentum, the more likely its direction is ``switched'' after yet another merger. 

However, host halo C8 seems to be the exception to this rule: just like C3 and C5 it is one of the few hosts with a very modest tendency for corotating satellites, but has an average spin parameter ($\lambda = 0.0231$). However we need to keep in mind that it is the youngest system under investigation. And its evolution gives us the key to understanding the missing signal: from a visual inspection of its mass accretion history and particle distribution over time we could infer that it just recently merged with two other haloes of nearly the same size, i.e. it is a (recent) triple merger! 
\citet{Vitvitska02} \citep[see also][]{Gardner01} found in their simulations that haloes with recent mergers (after $z=3$) generally show a larger spin parameter than ``quiet'' haloes. But even if the internal angular momentum of the resulting host halo has a ``stable'' direction, the satellites have been swirled around violently during the merging event and thus can be expected to have rather random orbit orientations, just as we found in the case of C8.

\section{``Observing'' the simulations} \label{sec:observer}
We have seen in our simulations, that subhaloes are preferentially moving on prograde orbits -- around 60\% of the subhaloes are corotating with the host halo. Could this signal be observed? Assuming that all our satellite haloes carry enough baryonic matter, in particular stars and gas, to be detectable by an observer, we are facing the question how the systems in our simulations appear from the observer's point of view.

\subsection{Cumulative line-of-sight velocity distribution}
\begin{figure*}
\begin{center}
	\epsfig{file=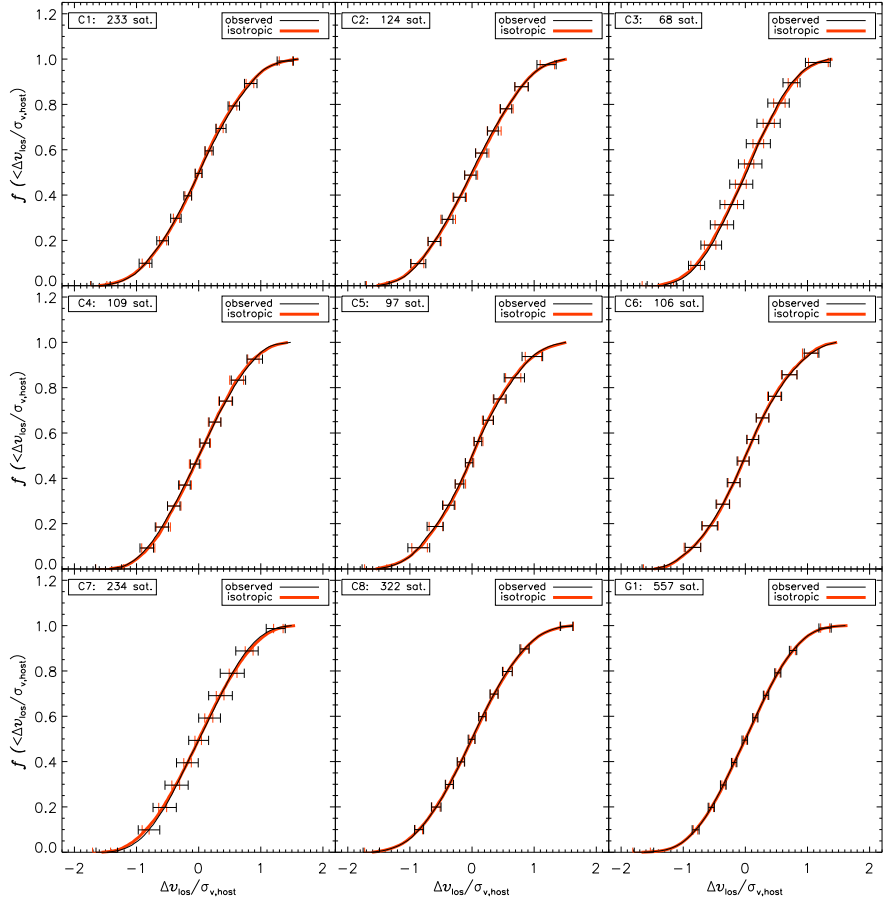, width=0.7\textwidth, angle=0}
\end{center}
\caption
{The cumulative distribution of line-of-sight velocities, normalized to the hosts' velocity dispersion
$\sigma_{v,\rm host}$ and averaged over the results from one hundred random lines of sight. The thick red
lines are ideal distributions for isotropic directions of velocities (but the same absolute
values). Error bars represent the standard deviation for a few sample points. There is practically no difference between the isotropic and ``observed'' curves, indicating an isotropic distribution of line-of-sight velocities.}
\label{f:radvelo}
\end{figure*}

\noindent
Usually an observer is not in the fortunate situation to know the positions and velocities of the satellites in all three dimensions. Yet it is possible to determine the velocity of a satellite along the line of sight and its projected position relative to the (density) centre of the host halo. In order to get a first impression of the velocity distribution of satellites, we looked at the cumulative distribution of line-of-sight velocities with respect to the host halo. We chose one hundred (isotropically distributed) random lines of sight and then averaged over the individual 1D velocity distributions. The resulting mean line-of-sight velocity distributions (with respect to the rest frame of the host halo and normalized to the velocity dispersion $\sigma_{v,\rm host}$) are shown in \Fig{f:radvelo}. The error bars represent the standard deviation.

If the velocities of the satellites were isotropically distributed (i.e. random directions with respect to the line-of-sight, but the same absolute values), we would observe the thick red lines in \Fig{f:radvelo}. Our data curves (thin black lines) are coinciding nearly perfectly with their corresponding isotropic curves. A KS test simply confirms the eye-balled result: the probability of the data curves being consistent with the isotropic curves is practically 100\%. This is actually not too surprising as the signal for a majority of prograde orbiting satellites is rather weak, but we could have expected a scattering in the data curves introduced by observing the halo from various viewing angles. However, the standard deviations for data and isotropic curves are nearly identical and rather small. The line-of-sight velocities are isotropically distributed, and hence we need to devise a different approach to observationally confirm the existence of prograde orbits.

%

\subsection{Classifying prograde and retrograde orbits}\label{class-pro-retro}

\begin{figure}
\begin{center}
	\epsfig{file=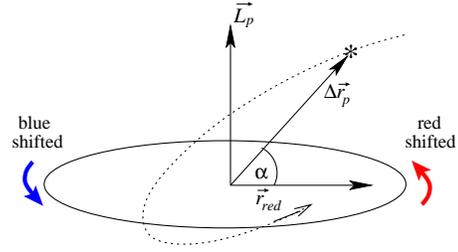, width=6cm, angle=0}
\end{center}
\caption
{Classifying observed satellites into prograde and retrograde orbits: The figure shows the projection plane perpendicular to the line of sight. $\vec{L}_p$ is the projected angular momentum of the host halo, $\Delta\vec{r}_p$ is the projected distance between the host halo centre and a satellite (marked with an asterisk). The angle $\alpha$ between this distance vector and the vector from the host halo centre to the red end of the host ($\vec{r}_{\rm red}$) is used for deciding whether the satellite is located on the red end or the blue end side of the host halo.}
\label{f:red_end_method}
\end{figure}

\begin{figure*}
\begin{center}
	\epsfig{file=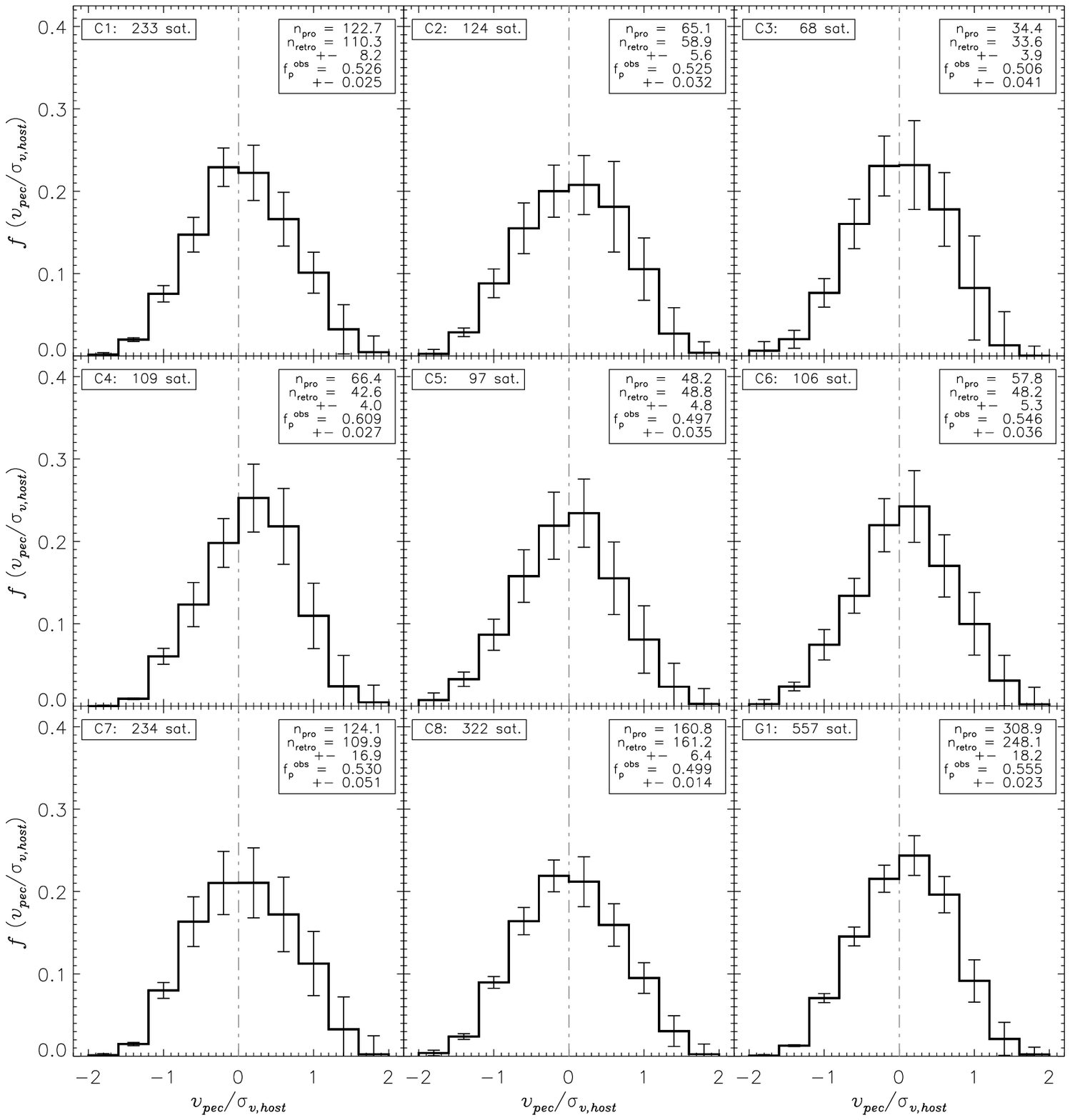, height=0.7\textwidth, angle=0}
\end{center}
\caption
{The peculiar velocity distribution (normalized to the hosts' velocity dispersion $\sigma_{v,\rm host}$) for all nine simulations. Here, velocities are defined to be positive, if the observer classifies the satellite as prograde, negative velocities mean retrograde motion.}
\label{f:obsvelo}
\end{figure*}

In what follows we lay out a method allowing an observer to distinguish prograde and retrograde orbiting satellites (cf. \Fig{f:red_end_method}) which has already been used by \citet{Azzaro05} in a similar analysis (see also e.g. \citet{Zaritsky97}, for an observational study): One first decides whether the satellite is moving away from or towards the host by determining the line-of-sight velocity $\Delta v_{\rm los}$ with respect to the host halo. This needs to be compared with the host's sense of rotation: We call the approaching side of the host the ``blue end'', while the receding part is called ``red end'' and define $S=1$ and $S=-1$, respectively. Based upon this information we can now define a (one-dimensional) peculiar velocity for each satellite

\begin{equation} \label{Vpec}
 v_{\rm pec} = S \cdot \Delta v_{\rm los} \ ,
\end{equation}

\noindent
and hence a satellite is on a prograde orbit for \vpec$>0$ and retrograde for \vpec$<0$.

\subsection{Peculiar velocity distributions}
In \Fig{f:obsvelo} we present the results for the (differential) peculiar velocity distribution. We used again one hundred random lines of sight and the velocities are normalized by the hosts' velocity dispersion. All satellites to the left of the vertical dash-dotted ``zero'' line have been classified as retrograde, while all satellites to the right are observed to be prograde. In addition, the mean number of prograde and retrograde orbits (along with the $1\sigma$ deviation) is given for every host halo in the upper right corner of every panel. The signal of an excess of prograde orbits as found when using the full three-dimensional (velocity) information is confirmed, but the signal is weakened in comparison to our previous findings. We only get a mean observed prograde fraction of 53.3\% ($\pm$3.3\%) as opposed to 59.0\% from \Sec{sec:KStest}. The individual observed prograde fractions $f_p^{\rm obs}$ for each dark matter host are given in the inset panels of the respective plot of \Fig{f:obsvelo}.

\begin{figure}
\begin{center}
	\epsfig{file=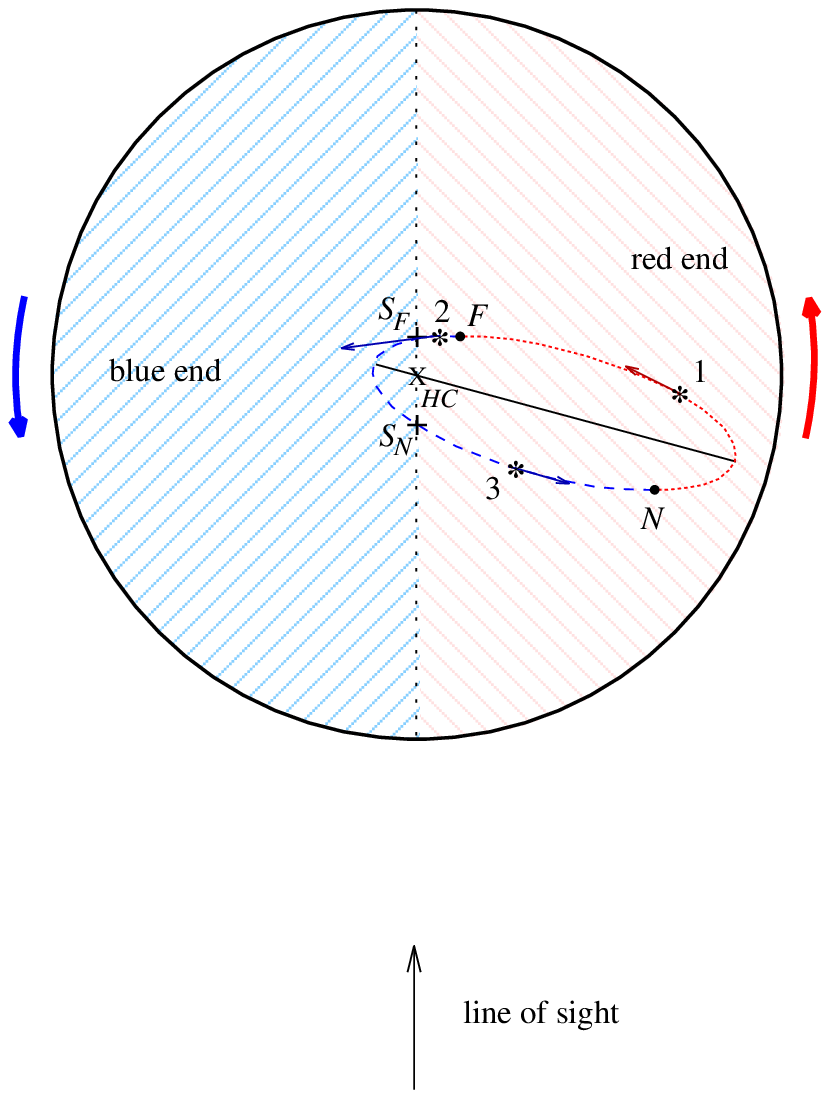, angle=0, width=5.3cm} 
\end{center}
\caption{The host halo and one of its satellites on an idealised elliptic orbit as seen from ``above''. The observer is somewhere beyond the bottom of the page, looking at this system along the line of sight. $HC$ marks the host halo centre, $F$ is the farthest and $N$ is the nearest point of the orbit relative to the observer. The part of the orbit when the satellite is approaching is dashed (blue), the receding part is dotted (red). Intersections of the orbit with the imaginary line dividing the host halo into red and blue end are named $S_F$ and $S_N$. The satellite actually is on a prograde orbit, but will be misclassified as retrograde, if it is observed at, for instance, position 2 or 3 (between F and $S_F$ or $S_N$ and $N$, respectively.)}

\label{f:red_end_orbit}
\end{figure}

\subsection{Discussion of the ``observer's'' result}

\subsubsection{Misclassification of orbits} \label{sec:misclassification}
The weakening of the prograde fraction for the observer is the result of a possible false classification of the satellites' sense of rotation. The method applied (i.e. \Eq{Vpec}) is based upon one not yet mentioned assumption: it will only work
faultlessly,
if the satellites are on circular orbits, which usually is not the case. Previous studies have shown that satellites can be on highly eccentric orbits, and their eccentricity distributions can be fitted by a Gaussian with a mean eccentricity of $e_0=0.61$ (\cite{Stuart2}, but see also \cite{Tormen97,TormenDiaferioSyer98,Ghigna-ea98}).

In observational studies \citep[e.g.][]{Zaritsky97, Carignan-ea97, Azzaro05} this method is used without mentioning its limitations, thus we here explain the cause of misclassification in more detail.

\Fig{f:red_end_orbit} elucidates the situation by viewing a sample satellite-host system from  ``above'', i.e. perpendicular to the line-of-sight, which lies in the plane of the paper. 
In compliance with our terminology, the part of the host halo which is hashed with red lines is called ``red end'' -- it is the part of the host halo that is receding from us. The ``blue end'' is defined accordingly. The ellipse represents a (idealised) part of an eccentric satellite orbit whose measured relative line-of-sight velocity $\Delta v_{\rm los}$ is positive in the red dotted part and negative in the blue dashed part. The small velocity vectors attached to the $*$-symbol for the satellite at the points 1, 2 and 3 should indicate that the satellite is corotating with the host halo and hence on a prograde orbit. However, we can readily see in \Fig{f:red_end_orbit}, that the satellite will be misclassified, if it happens to be observed in the blue (dashed) region of its orbit but on the red end side of the host halo (cf. positions 2 and 3). Even though the satellite shares the sense of rotation with the host, it appears to be counter-rotating; it approaches us while the host is receding. Such situations can occur when the turning points of the orbit, i.e. the points where the line-of-sight velocity of the satellite changes its sign ($F$ is the farthest and $N$ is the nearest point to the observer), are not aligned with the line of sight.

\begin{figure}
\begin{center}
	\epsfig{file=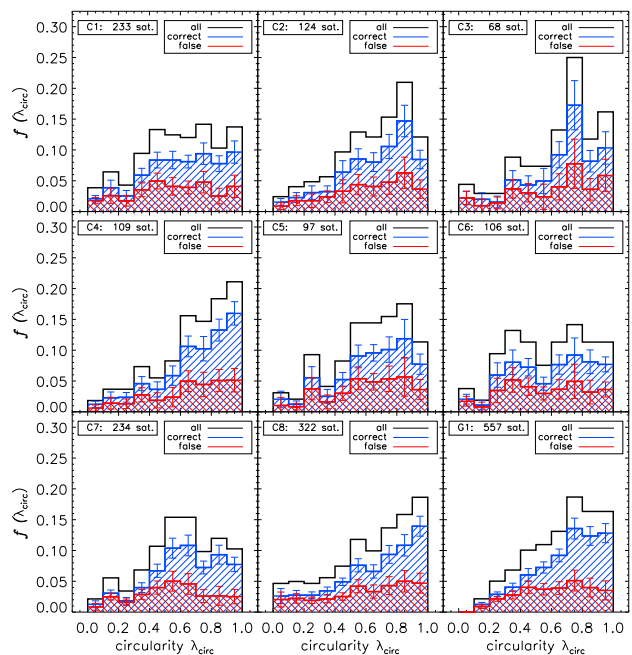, angle=0, height=0.53\textwidth}  
\end{center}
\caption{Circularity distribution of satellites, i.e. the fraction of satellites as a function of circularity $\lambda_{\rm circ}$ divided by the size of each bin (solid lines). For circular orbits we have a higher fraction of correctly classified satellites (blue hashed area) than misclassified satellites (red hashed area).}
\label{f:circularity}
\end{figure}

\begin{table*}
\begin{center}
\begin{tabular}{ccccc}
\hline
host & $f_p^{}$ & $c$ & $f_p^{\rm exp}$ & $f_p^{\rm obs}$\\
\hline\hline
C1 & 0.575 & $0.665\pm0.060$ & $0.525\pm0.009$ & $0.526\pm0.025$ \\
C2 & 0.581 & $0.656\pm0.077$ & $0.525\pm0.012$ & $0.525\pm0.032$ \\
C3 & 0.544 & $0.665\pm0.091$ & $0.515\pm0.008$ & $0.506\pm0.041$ \\
C4 & 0.706 & $0.684\pm0.067$ & $0.576\pm0.028$ & $0.609\pm0.027$ \\
C5 & 0.474 & $0.666\pm0.081$ & $0.491\pm0.004$ & $0.497\pm0.035$ \\
C6 & 0.689 & $0.638\pm0.075$ & $0.552\pm0.028$ & $0.546\pm0.036$ \\
C7 & 0.620 & $0.679\pm0.082$ & $0.543\pm0.020$ & $0.530\pm0.051$ \\
C8 & 0.531 & $0.676\pm0.060$ & $0.511\pm0.004$ & $0.499\pm0.014$ \\[1ex]
G1 & 0.594 & $0.699\pm0.058$ & $0.538\pm0.011$ & $0.555\pm0.023$ \\ \hline\hline

$<\cdot>$ & 0.590 & $0.670\pm0.016$ & $0.531\pm0.024$ & $0.533\pm 0.033$ \\\hline\hline

\end{tabular}
\caption{A summary of prograde fractions: $f_p$ is the fraction derived using the 3D information provided by the simulations (cf. \Table{t:kstestnum}), $f_p^{\rm exp}$ the expected observed fraction of prograde satellites based upon \Eq{eq:proratio} and $f_p^{\rm exp}$ the actual observed fraction (cf. \Fig{f:obsvelo}). $c$ is the mean fraction of correctly classified orbits (from one hundred lines of sight). Mean values, averaged over all simulations, are given again in the last line.}
\label{t:obsprofrac}
\end{center}
\end{table*}

\subsubsection{Origin of misclassification}
In order to support our theory of misclassification, we consider its sources in more detail. We are in the unique situation where we can distinguish between ``correctly'' and ``falsely classified'' orbits as we can compare the observer's classification to the three-dimensional classification used in \Sec{sec:simus}.  

We argued that the non-circularity of the orbits is the (possible) origin for a misclassification. In order to verify this conjecture, we use the circularity parameter $\lambda_{\rm circ}$, defined as the ratio of the angular momentum of the satellite orbit $L_{\rm sat}$ to the angular momentum of a corresponding circular orbit $L_{\rm circ}$ with the \emph{same} orbital energy $E_{\rm sat}$ \citep[cf.][]{Stuart2}:

\begin{equation}
\lambda_{\rm circ} = \frac{L_{\rm sat}(E_{\rm sat})}{L_{\rm circ}(E_{\rm sat})}
\end{equation}

\noindent 
\Fig{f:circularity} shows the binned circularity distribution for all satellites (black line), correctly classified satellites (blue hashed area) and falsely classified satellites (red hashed area). Satellites with circularity $\lambda_{\rm circ}=1$ are on circular orbits whereas radial orbits will have $\lambda_{\rm circ}=0$. However, a circularity of, for instance, $\lambda_{\rm circ}=0.9$ corresponds to an eccentricity of~$0.6$ \citep[cf. Fig.11 in][]{Stuart2} and hence small deviations from $\lambda_{\rm circ}=1$ already imply rather eccentric orbits! \Fig{f:circularity} hence proves that satellites with a high circularity are more likely to be correctly classified than their corresponding low circularity counterparts, stressing that the deviation from circular orbits increases the probability for misclassification.

\subsubsection{Quantitatively understanding misclassifications}
In Appendix~\ref{app:obsprofrac}, we show how the observed prograde fraction will be weakened due to misclassification, assuming that there is an equal probability of wrong classification for both pro- and retrograde orbits (the variation in our simulations is at most a few percent).
This follows from the simple consideration that all wrongly classified prograde orbits will actually be categorized as retrograde and vice versa. If there is, for instance, a majority of prograde satellites, more satellites will ``move'' from prograde to retrograde than from retrograde to prograde and hence the prograde fraction is weakened. We yield the following equation for the observed ratio of prograde $n_p$ to retrograde orbits $n_r$:
 
\bq
\label{eq:proratio}
q^{\rm obs} = \frac{n_p^{\rm obs}}{n_r^{\rm obs}} = \frac{c\cdot q + (1-c)}{c + q\cdot (1-c)}
\eq	
with $q = n_p/n_r$ being the ratio of prograde to retrograde 
orbits in the 3D-simulation and 

\bq
c = \frac{n_p^{\rm right}}{n_p} = \frac{n_r^{\rm right}}{n_r} 
\eq

\noindent
being the probability of right classification.

A majority of prograde orbits in our simulations can only change into a minority for an observer, if $c$ becomes less than 50\%, provided that our statistical sample is big enough. We determined the mean fraction of correct classification $c$ for every simulation by averaging over hundred randomly chosen lines of sight. We found $c$ to be around 67\%, the individual values are listed in \Table{t:obsprofrac}. From this value for $c$ and the ratio $q$ taken from our 3D-simulations, we can now use \Eq{eq:proratio} to deduce an \textit{expected} observed prograde fraction $f_p^{\rm exp}$. These predicted prograde fractions are given in \Table{t:obsprofrac} and can be compared to the actually observed fractions $f_p^{\rm obs}$ (cf. \Fig{f:obsvelo}). We find that our anticipated observed prograde fractions fit very well the mean observed prograde fractions. The mean value for the observed prograde fraction of all simulations was found to be 53.3\%, which is close to the predicted value of 53.1\%.

\begin{figure}
\begin{center}
	\epsfig{file=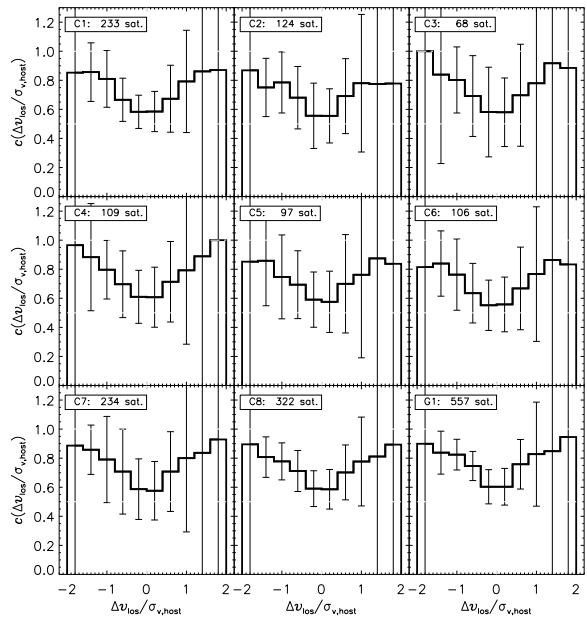, angle=0, height=0.5\textwidth} 
 
\end{center}
\caption{The fraction of correctly classified satellites as a function of relative line-of-sight velocity $\Delta v_{\rm los}$. The probability for correct classification is higher for high velocities. The error bars are the standard deviation from averaging the results over one hundred random lines of sight.}
\label{f:obsvelodist}
\end{figure}

\subsubsection{Overcoming misclassification}
\label{sec:overcoming}
The observer usually does not know the circularity of the orbit and also has no other means to check whether the satellite is residing in a disadvantageous region or not. However, he will be able to determine, if the satellite is very close to the border of such a region: close to the turning points (the nearest or farthest orbit point, $N$ or $F$ in \Fig{f:red_end_orbit}, respectively) its line-of-sight velocity $\Delta v_{\rm los}$ is very small. Thus, we suspect that a small line-of-sight velocity goes along with an increased probability of misclassification (but still less than 50\%).

To examine this more closely, we plot the fraction of correctly classified satellites per line-of-sight velocity bin in \Fig{f:obsvelodist}. The error bars give the standard deviation and are quite big for two reasons: first, the size of the disadvantageous region of an orbit with given eccentricity (or circularity) can change strongly with the direction of the line of sight and secondly, there are only a few satellites with high velocities, so there result large deviations in the probabilities in the corresponding bins.

Nevertheless we find the trend sought after:
For low velocities $|\Delta v_{\rm los}| \approx 0$, the correct fraction is of the order of 60\%, whereas for high velocities we nearly gain a 100\% probability of correct classification. An observer could now make use of the different probabilities of misclassification by weighing the observed data with an appropriate function giving high-velocity data more weight than low-velocity data. We applied a simple $\Delta v_{\rm los}$-weighing to the data used for the preparation of \Fig{f:obsvelo} and when averaging the newly derived prograde fractions over all nine simulations, we now gain a mean of 54.4\%. There is still a discrepancy due to the non-circularity of the orbits, but the original signal is enhanced and closer to the ``real'' 3D signal.

A similar dependency can be observed for subhaloes with small projected distances from their host halo centre:
These also are more probable to be misclassified. We examined this effect by just neglecting all satellites with a projected distance smaller than a specified fraction of the virial radius of its host. For higher minimum distances we yielded marginally growing observed prograde fractions, up to 54.0\% when neglecting all satellites within 40\% of the virial radius of the corresponding host.

\subsection{The effect of interlopers} \label{sec:interlopers}

Observationally it is quite difficult to determine whether a satellite belongs to the host halo or just happens to lie in front of (or behind) the host along the line of sight. To model these effects and quantify the influence of such interloping subhaloes, we included all objects in the 2D analysis that lie within the cylindrical tube centred about the position of the host halo and with radius \Rvir. We only consider satellites out to a maximum distance of 5\,\Rvir\ (cf. \Sec{sec:environment}) as possible sources of contamination.

In \Table{t:interloper}, we summarize the resulting (observed) prograde fraction when including interlopers. These numbers are accompanied by the fractional increase of the number of satellites and the absolute difference to the situation where interlopers are neglected. The new mean fraction is hardly affected and only shows a marginal decrease of about 1\%. We therefore conclude that interlopers have an insignificant effect on the signal as measured by an observer.

\begin{table}
\begin{center}
\begin{tabular}{cccr}
\hline
host & $n_{\rm int}/n_{\rm sat}$ & $f_p^{\rm obs}$ & $\Delta f_p^{\rm obs}$\\ \hline\hline
C1 & 1.41 & 0.521 & $-0.005$ \\
C2 & 1.68 & 0.521 & $-0.004$ \\
C3 & 1.40 & 0.516 & $0.010$ \\
C4 & 1.33 & 0.601 & $-0.008$ \\
C5 & 1.32 & 0.498 & $0.001$ \\
C6 & 1.53 & 0.531 & $-0.015$ \\
C7 & 1.45 & 0.529 & $-0.001$ \\
C8 & 1.18 & 0.500 & $0.001$ \\[1ex]
G1 & 1.41 & 0.547 & $-0.008$ \\\hline\hline
   & \multicolumn{3}{c}{$<f_p>^{\rm obs}$ $0.529\pm0.029$}\\ \hline\hline
\end{tabular}
\caption{The interlopers and their effect on prograde fractions. $n_{\rm int}$ is the number of satellites including interlopers, $\Delta f_p^{\rm obs}$ is the absolute difference between the observed prograde fraction including interlopers and the previously found value.}
\label{t:interloper}
\end{center}
\end{table}

\subsection{Alignment of disk and dark matter halo angular momentum}
The problem still remains that an observer will only be able to measure the distribution and motion of visible matter, not the dark matter haloes themselves. Thus, the question arises, whether the observed trend of prograde fraction still holds for visible matter. The observer determines the rotation axis of a disk galaxy by measuring red- and blueshifts in the disk material. We may safely assume that the rotation axis of the disk is well aligned with its own angular momentum, but there are doubts about the alignment of disk and dark matter halo angular momentum. For example, the simple picture of disk formation by \citet{FE80} leads to an alignment of disk and halo angular momentum, but e.g. \citet{Bailin-ea05} found that the disk rotation axis and the minor axis of the dark matter halo coincide very well within $0.1\,R_{\rm vir}$, probably due to their joint evolution. However, the latter study also emphasizes that the presence of baryonic matter changes the orientation of the dark matter halo and in \citet{BailinSteinmetz05} a typical misalignment between the angular momentum vector and the minor axis of 25$^\circ$ was found. However, since we have seen in our analysis that a change of the host angular momentum only had effects of typically a few percent in the prograde fraction, we dare to say that the influence of a disk should not change the result in principle.

\section{Discussion}\label{sec:discussion}
In this Section we are going to discuss our results and compare them to the findings (and observations) of other groups, respectively.

While we analyzed our 3D data by determining the sense of rotation for each subhalo, counting subhaloes on prograde orbits and thus determining a prograde fraction of 59\%,
a different method for investigating the direction of rotation of subhaloes was used by \cite{Shaw05}. They compared the angular momentum of \emph{particles} in subhaloes (rather than subhaloes themselves) to the host's angular momentum and found a strong preference of prograde motion (cf. Fig. 31 in their study). This method agrees more or less with the utilization of $\cos\Phi_{\rm net}$, the net orbital angular momentum of all satellites as presented in \Sec{sec:KStest}. However, applying this method to our data, we also obtain a rather large fraction of $\approx 89\%$ corotating (satellite) particles. 

Our results are further consistent with the theoretical signal found by \citet{Azzaro05}, who determined the observed prograde fraction for one dark matter halo drawn from a cosmological simulation. They used the same technique to project the simulation data into the observer's plane, but additionally mapped magnitudes onto their dark matter subhaloes using the (semi-analytical) prescription of \citet{Kravtsov04}. They found an observed prograde fraction of around 55--60\%, depending on the mass (or magnitude, correspondingly) of the satellites: massive satellites were found to reside preferentially on prograde orbits. We could not detect such a mass dependence in our simulations, as our mass-weighted curves did not necessarily enhance the signal. However, their minimum signal of 55\% prograde satellites (when including also low-mass satellites, see their Table 2) agrees well with our result of 53\%.

\citet{Azzaro05} further analyzed a set of observational data from the SDSS catalogue and actually found a strong signal of 61\% prograde satellite galaxies. This is in contrast to our results: we merely predict a fraction of about 53\% corotating subhaloes which actually poses a challenge for an observer to detect. We could imagine several reasons for this discrepancy, whose (tiny) individual effects may add up and thus bring our findings into agreement with the \citet{Azzaro05} results:

\begin{itemize}
\item \emph{Determining the mean prograde fraction.} Using the observational data, \citet{Azzaro05} selected 43 similar host galaxies and considered the total sample of 76 satellite galaxies to belong to a single, fictitious primary. In contrast to this, we calculated prograde fractions of subhaloes for every individual host halo, before averaging the result. When adhering to the same approach of stacking the data prior to determining the respective fractions, we end up with 53.9\% prograde orbits. Grouping together only satellites of host haloes which are of roughly the same size and mass (C2, C3, C4, C5, C6) leads to the same result.

\item \emph{Missing baryonic matter in subhaloes.} Most probably not all of our subhaloes would carry baryonic matter, some of them would therefore escape detection by an observer. It is likely that such dark haloes have a low total mass or low central density. If we thus restrict our subhaloes and introduce a larger minimum mass, minimum density or a lower threshold for the maximum value of the rotation curve ($V_{\rm max}$, similar to observations), we might get a better impression of the true satellite galaxy distribution and orientation. We experimented with several threshold values, but were unable to obtain mean prograde fractions above 54\%.

\item \emph{Neglecting satellites with a small projected distance.} Since it is intrinsically difficult to clearly detect satellite galaxies which are very close to the centre, \citet{Azzaro05} neglected all satellites with a projected distance smaller than 20 kpc. 
This can be advantageous, since satellites close to the centre (in projection) are more likely to be misclassified than satellite galaxies residing at farther distances (cf. \Sec{sec:overcoming}). Thus, the effect of misclassification might decrease, when neglecting such suhaloes with small projected distances to the centre of their respective host. We found in our simulations only a minor improvement of the signal by at most 1\%.

\item \emph{Number statistics.} The observational sample of \citet{Azzaro05} consists of only 76 satellites, whereas our total sample contains 1850 subhaloes. Thus, the difference in the prograde fraction might still be a statistical variance due to low numbers.
\end{itemize}
 
In summary, we are unable to reproduce such a high value of 61\% observed prograde orbits as found by \citet{Azzaro05}. However, \citet{Zaritsky97} presented in their observational study 115 satellite galaxies belonging to 69 host galaxies (isolated spiral galaxies). 
Orbit orientations were determined for 95 of the satellites with the method described in \Sec{class-pro-retro}. They found 49 satellites on prograde orbits, yielding a prograde fraction of 51.6\%, which is perfectly consistent with our prediction of an observed prograde fraction.

\citet{Carignan-ea97} studied the disk galaxy NGC 5084 with 8 satellites, of which all except one were found to be on a retrograde orbit. This neither supports the prediction of \citet{Azzaro05} nor supports our results. An excess of retrograde orbits may be explained by the fact that dynamical friction is enhanced for satellites orbiting in the same direction as the rotation of the host galaxy. This effect was studied by \citet{QuinnGoodman86} using $N$-body simulations, who found the lifetime of retrograde satellites considerably larger. However, \citet{Penarrubia02} found the effect to give only an increase of a few per cent in lifetime, depending on the triaxiality of the host galaxy halo, so this effect alone could not explain the majority of retrograde satellites in \citet{Carignan-ea97}. However, since the number of satellites considered in that study is rather low, we consider this signal statistically insignificant.

\section{Conclusion} \label{sec:conclusions}

We investigated the sense of orientation of satellite orbits in cosmological dark matter haloes at redshift $z=0$. Our set of simulations consisted of eight cluster-sized haloes and one simulation of a Milky Way-type galactic dark matter halo.

From the theory of the generation of angular momentum in dark matter haloes, we expected to get a trend towards satellites corotating with the host. 

We found a mean majority of 59\% prograde orbits averaged over all nine simulations. Three of our simulations expressed a rather isotropic distribution of orientations of orbits with respect to the host. Two of them could be assigned a rather small spin parameter of the host, while the third could be confirmed as a recent (triple) merger leading to random orientations of satellite orbits.

We validated our results by carrying out a Kolmogorov-Smirnov test for the cumulative distribution of angles between satellite orbit angular momentum and host halo spin. All simulations, except the already mentioned exceptions, 
are clearly not in agreement with an isotropic distribution. There is also no sign for a symmetry between prograde and retrograde orbits in these simulations, stressing that the excess of corotating satellites is a reliable prediction of the hierarchical \LCDM\ structure formation scenario.

We chose to use the inner host angular momentum, i.e. within 30\% of the virial mass $M_{\rm vir}$ (corresponding to $\approx 0.2\,R_{\rm vir}$) for defining the rotation of the host halo. However, other choices (at 50\%, 100\% of $M_{\rm vir}$) lead to the same mean prograde fraction of 59\%. Thus, uncertainties in the angular momentum only have a minor effect.

We further checked the influence of the environment outside the virial radii $R_{\rm vir}$ of the hosts by including satellites up to $5\,R_{\rm vir}$. Even though the prograde fraction decreased and the orientations became more isotropic, the signal yet remained -- in agreement with the picture that the environment of matter and satellites determines the sense of host halo rotation.

Thus, we conclude that there is a trend towards an overbalance of corotating subhaloes in $\Lambda$CDM simulations at redshift $z=0$, regardless of the inner host boundary defining the host angular momentum and the outer boundary determining the sphere within which satellites are included. We could not confirm any relation between triaxiality and prograde fraction and, even more important, we find our signal in the cluster-sized simulations as well as in the galaxy-sized simulation.

The question now remained, how feasible it is to observationally test the prediction of having roughly 60\% satellites on orbits corotating with their host halo. To obtain an answer, we chose 100 random lines of sight projecting our three-dimensional data into an observer's plane. 
The signal still remained, yet weakened to $53\% \pm 3\%$, which is difficult to be detected.

The origin of this weakening lies in the observer's method for distinguishing between pro- and retrograde orbiting satellites by comparing their motion with the part of the host where they reside (cf. \Sec{class-pro-retro}). This commonly used method works perfectly for circular orbits, yet satellite orbits are not necessarily circular but rather exhibit a distribution of circularities \citep[cf.][]{Stuart2}.
We showed that the recovery of the original signal can be improved by weighing the satellites with their relative line-of-sight velocities, yielding an observed prograde fraction of 54\%.

Additionally, we studied the effect of interlopers, satellites which pass by the observed cluster or galaxy along the line of sight, and found it to be negligible (i.e. less than 1\%).

Comparison with observational work yields different results: the study of \citet{Zaritsky97} agrees with our findings, \citet{Azzaro05}'s observational results give a much higher prograde fraction, whereas \citet{Carignan-ea97} found for NGC 5084 a strong excess of retrograde satellite galaxies.
More galaxy-sized simulations, maybe even including baryonic matter, would be necessary in order to study the prograde fractions on smaller scales in more detail, which would then enable a better comparison with observational results.




\section*{Acknowledgments}
We thank Anatoly Klypin for providing the galactic halo simulation as well as some stimulating discussions. We further benefited from valuable debates with Stuart Gill, Chris Power, Brad Gibson, Stefan Gottl\"{o}ber and Volker M\"{u}ller. We acknowledge funding through the Emmy Noether Programme by the DFG (KN 755/1). The simulations of the eight cluster-sized haloes presented in this paper were carried out on the Beowulf cluster at the Centre for Astrophysics~\& Supercomputing, Swinburne University. The galactic halo simulation has been performed utilizing the AIP's Sanssouci Dual-Opteron cluster.

\appendix

\section{Weakening of the observed prograde fraction}
\label{app:obsprofrac}

Assuming that the fraction of correctly classified satellites is the same for satellites on prograde and retrograde orbits, we can make following considerations in order to estimate the ``observed'' prograde fraction for our simulations.

Let $n_p$ be the number of prograde and $n_r$ be the number of retrograde orbits in one of our 3D-simulations. An observer of the satellites in this simulation may be mistaken in classifying some of the satellites due to problems referred to in \Sec{sec:misclassification}. Thus, the number of prograde orbits $n_p^{\rm obs}$ will be the sum of correctly classified prograde orbits $n_p^{\rm right}$ and satellites, which are actually retrograde, but have been wrongly classified ($n_p^{\rm false}$):
\bq
\label{eq:obsprograde}
n_p^{\rm obs} = n_p^{\rm right} + n_p^{\rm false} \ .
\eq
Equivalently, the number of retrograde orbits $n_r^{\rm obs}$ is the sum of correctly determined retrograde orbits $n_r^{\rm right}$ and wrongly classified retrograde orbits $n_r^{\rm false}$:
\bq
n_r^{\rm obs} = n_r^{\rm right} + n_r^{\rm false}
\eq
Since we know that the latter orbits can only be prograde orbits in truth and vice versa, we can write:
\bq
\label{eq:false}
\begin{array}{ll}
n_p^{\rm false} &= n_r - n_r^{\rm right}\\
\\
n_r^{\rm false} &= n_p - n_p^{\rm right}\quad. \\
\end{array}
\eq

Let us now assume that the probability of correct classification $c$ is constant, i.e. independent of the
sense of rotation:
\bq
\label{eq:corrclass}
	c: = \frac{n_p^{\rm right}}{n_p} = \frac{n_r^{\rm right}}{n_r} \quad .
\eq
Equation (\ref{eq:obsprograde}) can then be re-written using equations (\ref{eq:false}), (\ref{eq:corrclass}) and we obtain:
\bq
\begin{array}{ll}
	n_p^{\rm obs} 	&= n_p^{\rm right} + n_p^{\rm false}\\
			&= n_p^{\rm right} + (n_r - n_r^{\rm right})\\
			&= c\cdot n_p + (n_r - c\cdot n_r)\\
			&= c\cdot n_p + (1-c) \cdot n_r\\
\end{array}
\eq
For $q$ representing the constant ratio of prograde to retrograde orbits in our 3D-simulations,
\bq
q:=\frac{n_p}{n_r} \quad ,
\eq
we can express the number of observed prograde orbits by
\bq
\begin{array}{ll}
n_p^{\rm obs} &= c \cdot q\cdot n_r + (1-c) \cdot n_r\\
	  &= [\,c\cdot q + (1-c) ] \cdot n_r\ .\\
\end{array}
\eq
Similarly, we yield for the observed number of retrograde orbits:
\bq
\begin{array}{ll}
n_r^{\rm obs} 	&= c\cdot n_r + (1-c) \cdot n_p\\
		&= c\cdot n_r + (1-c) \cdot q \cdot n_r\\
		&= [\,c+ q\cdot(1-c)] \cdot n_r\ .\\
\end{array}
\eq
The ratio of observed prograde to observed retrograde orbits $q^{\rm obs}$ is then:
\bq
q^{\rm obs} = \left(\frac{n_p}{n_r}\right)^{\rm obs} = \frac{c\cdot q + (1-c)}{c + q\cdot (1-c)}
\eq
which is \emph{not} simply the same as $q$!

We conclude that the observed ratio of prograde to retrograde orbits is only equal to the real ratio $q$, if $c=1$ or $c=0$, meaning all or none satellites are classified correctly. Note, that the case of 100\% prograde ($q=1$) or 100\% retrograde satellites ($q = 0$) must not be allowed, because then our assumption of a constant probability of correct classification $c$ fails -- it will not be well defined anymore.\\

%
%
%


\bsp

\label{lastpage}

\end{document}